\documentclass[10pt,preprint]{emulateapj}
\usepackage{apjfonts}
\newcommand{\sol}{\odot}
\newcommand{\del}{\nabla}
\newcommand{\cross}{\times}
\newcommand{\avg}{\bar}
\newcommand{\scrD}{\mathcal{D}}
\newcommand{\scrR}{\mathcal{R}}
\newcommand{\ms}{$\mathrm{m}\thinspace \mathrm{s}^{-1}$}

\shorttitle{Rapid Rotation and Modulated Convection}
\shortauthors{Brown, Browning, Brun, Miesch \& Toomre}

\begin{document}

  \slugcomment{Submitted to ApJ}

  \received{April 30, 2008}
  \revised{July 1, 2008}
  \accepted{August 12, 2008}

  \title{Rapidly Rotating Suns and Active Nests of Convection}
  \author{Benjamin P.\ Brown}
  \affil{JILA and Dept.\ Astrophysical \& Planetary Sciences, University of Colorado, Boulder, CO 80309-0440}
  \email{bpbrown@solarz.colorado.edu}
  \author{Matthew K.\ Browning\altaffilmark{1}}
  \affil{Dept.\ of Astronomy, University of California, Berkeley, CA 94720-3411}
  \altaffiltext{1}{present address: Dept of Astronomy \& Astrophysics, U.~Chicago, 5640~S.~Ellis~Ave, Chicago, IL 60637}
  \author{Allan Sacha Brun}
  \affil{DSM/IRFU/SAp, CEA-Saclay \& UMR AIM, CEA-CNRS-Universit\'e Paris 7,
         91191 Gif-sur-Yvette, France}
  \author{Mark S.\ Miesch}
  \affil{High Altitude Observatory, NCAR, Boulder, CO 80307-3000}
  \and
  \author{Juri Toomre}
  \affil{JILA and Dept.\ Astrophysical \& Planetary Sciences, University of Colorado, Boulder, CO 80309-0440}

  \begin{abstract}
    In the solar convection zone, rotation couples with intensely
    turbulent convection to drive a strong differential rotation and
    achieve complex magnetic dynamo action.  Our sun must
    have rotated more rapidly in its past, as is suggested by
    observations of many rapidly rotating young solar-type stars.
    Here we explore the effects of more rapid rotation on the
    global-scale patterns
    of convection in such stars and the flows of differential
    rotation and meridional circulation which are self-consistently
    established.  The convection in these systems is richly
    time dependent and in our most rapidly rotating suns a striking
    pattern of localized convection emerges.  Convection near the
    equator in these systems is dominated by one or two nests in
    longitude of locally enhanced convection, with quiescent streaming flow in
    between at the highest rotation rates.  These active nests of
    convection maintain a strong differential rotation despite their
    small size.  The structure of differential rotation is similar in
    all of our more rapidly rotating suns, with fast equators and
    slower poles.  We find that the total shear in differential rotation
    $\Delta \Omega$ grows with more rapid rotation while the relative
    shear $\Delta \Omega/ \Omega_0$ decreases.  In contrast, at more
    rapid rotation the meridional circulations decrease in energy
    and peak velocities and break into multiple cells of circulation
    in both radius and latitude.
  \end{abstract}
  \keywords{convection -- Sun:interior -- Sun:rotation --
    stars:interiors -- stars:rotation}


\section{Convection, Rotation and Magnetism}
Our sun is a magnetic star whose cycles of magnetic activity must
arise from organized dynamo action in its interior.
This dynamo action is achieved by turbulent plasma motions in the
solar convection zone, which spans the outer 29\% of the sun in
radius.  Here vigorous convective motions and rotation couple to drive
the differential rotation and meridional circulation.  
These global-scale flows are important ingredients in
stellar dynamo theory, providing shear which may build and
organize fields on global scales.
When our sun was younger, it must have rotated much more
rapidly, as is suggested both by the solar wind which continually removes
angular momentum from the sun and by many observations of rapidly
rotating solar-like stars.  In more rapidly rotating suns the
coupling between rotation and convection is strong and must continue to drive 
global scales of flow. Understanding the nature of convection, differential
rotation and meridional circulation in more rapidly rotating stars is
a crucial step towards understanding stellar dynamos.

The manner in which the sun achieves global-scale dynamo action is
gradually being sorted out.
Helioseismology, which uses acoustic oscillations 
to probe the radial structure of the star as well as the convective
flows beneath the surface, has revealed that the solar differential
rotation profile observed at the surface prints throughout the
convection zone with two prominent regions of radial shear.  The
near-surface shear layer occupies the outer 5\% of the sun,
whereas a tachocline of shear at the base of the convection zone
separates the strong differential rotation of that zone from the 
uniform rotation of the deeper radiative interior
\citep[e.g.,][]{Thompson_et_al_2003}.  The solar global magnetic dynamo,
responsible for the 22-year cycles of activity, is now believed to be
seated in that tachocline.  In such interface dynamo models, magnetic
fields generated in the bulk of the convection zone are pumped into
the tachocline where the radial shear builds strong toroidal
magnetic fields, with magnetic buoyancy leading to loops that rise
upward and erupt through the solar surface \citep[e.g.,]
[]{Charbonneau_2005,Miesch_2005}.  The differential rotation plays an
important role in building and organizing the global-scale fields
while the meridional circulations may be important for returning flux
to the base of the convection zone and advecting it equatorward, enabling cycles of magnetic
activity.  By studying the coupling of rotation and convection over a
range of conditions, we
are likely to learn about the operation of the current solar dynamo 
and about the nature of dynamos operating in our sun's past and in other 
solar-like stars.

Observations of young solar-like stars indicate that they rotate as
much as 50 times faster than the current solar rate.  Many of these more
rapidly rotating suns possess strong magnetic fields.
A correlation between rotation and magnetic activity is observed
over a range of stellar types and populations, indicating that more
rapidly rotating stars may have stronger stellar dynamos
\citep[e.g.,][]{Noyes_et_al_1984a,Patten&Simon_1996}.  Probing the nature of these
dynamos and the impact of faster rotation on the internal stellar
dynamics requires both accurate observations and detailed dynamical models of
the stellar interiors.  The faster flows of differential rotation are
much easier to detect than the relatively slow motions associated with
meridional circulations; observations across the HR diagram
indicate that differential rotation is a common feature in many stars.
Asteroseismic observations with the Kepler and Corot missions may soon begin to
constrain the internal rotation structure.  At present only
measurements of surface differential rotation are available, as assessed
with a variety of techniques including photometric
variability \citep{Donahue_et_al_1996, Walker_et_al_2007}, Doppler
imaging \citep{Donati_et_al_2003} and Fourier transform methods
\citep{Reiners&Schmitt_2003}. 

Advances in supercomputing have enabled 3-D
simulations that are beginning to capture many of
the dynamical elements of the solar convection zone.  Early
global-scale simulations of solar convection by
\citet{Gilman_1975,Gilman_1977,Gilman_1979} under the Boussinesq
approximation were extended by the pioneering work of
\citet{Gilman&Glatzmaier_1981}.  Such global-scale simulations of solar 
convection conducted in full spherical shells sought to capture the largest
scales of convective flows and began to study how they can establish
differential rotation and meridional circulations.  However, the range of
spatial and temporal scales present in solar convection are vast and
thus the computational resources required by the modeling are daunting.
Through recent advances in massively parallel computer architectures,
solar convection simulations are now beginning to make detailed contact with the
observational constraints provided by helioseismology \citep[e.g.,][]{
Brun&Toomre_2002, Miesch_et_al_2006, Miesch_et_al_2008}. Other efforts have focused on
the vigorous turbulence and the dynamo action achieved in the bulk of the 
solar convection zone \citep{Brun_et_al_2004}, with recent studies
beginning to include the tachocline as a region of penetrative
overshoot, shear, and magnetic field amplification \citep{Browning_et_al_2006}.  
Facilitated by these
computational advances, models of convection and dynamo action within the 
cores of A-type stars have also begun to be investigated \citep{Browning_et_al_2004, 
Brun_et_al_2005, Featherstone_et_al_2007}.

To date, most models of stellar differential rotation in stars like
our sun that rotate more rapidly have been carried out in 2-D under the simplifying
assumptions of mean field theory \citep[e.g.,][]{Rudiger_et_al_1998,
Kuker&Stix_2001, Kuker&Rudiger_2005_A&A, Kuker&Rudiger_2005_AN}.  The
time is ripe to pursue the question with fully 3-D simulations of
global-scale stellar convection.

In order to study solar-like stars that rotate faster,
our previous work focused on a series of 3-D compressible simulations within a spherical shell 
using the anelastic spherical harmonic (ASH) code for stars rotating
from one to five times the current solar rate \citep{Brown_et_al_2004}.  These preliminary 
hydrodynamic simulations explored how stellar convection changes with more rapid
rotation, including the differential rotation and meridional circulation that 
is achieved. Comparable studies have been carried out by \citet{Ballot_et_al_2007}
studying younger stars with deeper convection zones.  In our
simulations we found that remarkable
nests of vigorous convection emerge in the equatorial regions.  Namely, convective structures 
at low latitudes about the equator can exhibit strong spatial modulation with longitude, 
and at high rotation rates the convection is confined to narrow intervals (or nests) 
in longitude.  In the more turbulent simulations presented in this
paper, which span a larger range of rotation rates from one to ten times
the current solar rate, the phenomena 
of modulated convection has persisted and the active nests of
convection are prominent features at the higher rotation rates. 
These nests of localized convection persist for long intervals of time
and despite their small filling factor maintain a strong differential rotation.

The emergence of spatially localized convective states has been observed in other systems, 
particularly in theoretical studies of doubly-diffusive systems such as thermosolutal convection 
\citep[e.g.,][]{Spina_et_al_1998, Batiste_et_al_2006}, in laboratory studies of convection 
in binary fluids \citep[e.g.,][]{Surko_et_al_1991}, and in simulations of magnetoconvection
where isolated ``convectons'' have been observed \citep{Blanchflower_1999}.  
In shells of rapidly rotating fluid, temporally intermittent patches of localized
convection emerged in Boussinesq simulations of the geodynamo \citep{Grote&Busse_2000}.  
In many of these systems, spatial modulation occurs in the weakly nonlinear regime
close to the onset of convection.  In contrast, our simulations
of solar convection are in a regime of fully developed turbulent convection.

We describe briefly in \S\ref{sec:ASH} the 3-D anelastic spherical
shell model and the parameter space
explored by our simulations. In \S\ref{sec:convection} we discuss
the nature of convection realized in more rapidly rotating stars and the
emergence of spatially-localized patterns of convection.   In \S\S\ref{sec:global scale flows}-\ref{sec:energies} 
we examine the global-scale flows realized in our simulations, 
including differential rotation and meridional circulation, and their
scaling with more rapid rotation.  
A detailed exploration of the active nests of convection is presented in
\S\ref{sec:patches}.  We reflect 
in \S\ref{sec:conclusion} on the significance of our findings.


\section{Tools to Model 3-D Stellar Convection}
\label{sec:ASH}
To study through simulations the global scales of motion likely
achieved in stellar convection zones, we must employ a global model which
simultaneously captures the spherical shell geometry and admits the
possibility of zonal jets and large eddy vortices, and of convective plumes
that may span the depth of the convection zone.  The solar convection
zone is intensely turbulent and molecular values of viscosity and
thermal diffusivity in the sun are estimated to be very small.  As a
consequence, numerical simulations cannot hope to resolve all scales
of motion present in real solar convection and a compromise must be
struck between resolving the important dynamics within
small regions and capturing the connectivity and geometry of the
global scales. We opt here for the latter by studying a full shell of
convection.

\subsection{Anelastic Simulation Approach}
Our tool for exploring stellar convection is the anelastic spherical
harmonic (ASH) code, which is described in detail in
\cite{Clune_et_al_1999} and in \cite{Brun_et_al_2004}.  ASH is
designed to run efficiently on massively parallel architectures. ASH
solves the three-dimensional (3-D) anelastic equations of motion in a
rotating spherical shell using the pseudo-spectral method. The
thermodynamic variations and the three components of mass flux are
expanded in spherical harmonics to resolve their horizontal structure
and in Chebyshev polynomials to resolve their radial structure.  The
anelastic approximation is used to capture the effects of density
stratification without having to resolve sound waves which have short
periods (about 5 minutes) relative to the dynamical time scales of most
interest (months to years). The mass flux remains divergence-free by
using a poloidal-toroidal representation.  Temporal discretization is
accomplished using a semi-implicit Crank-Nicholson time-stepping
scheme for linear terms and an explicit Adams-Bashforth scheme for
nonlinear terms.  Under the anelastic approximation the thermodynamic
variables are linearized about their spherically symmetric and
evolving mean state with density $\bar{\rho}$, pressure $\bar{P}$,
temperature $\bar{T}$ and specific entropy $\bar{S}$.  Fluctuations
about this state are denoted as $\rho$, $P$, $T$ and $S$ respectively.  In the
uniformly rotating reference frame of the star, the resulting
hydrodynamic equations expressing the conservation of mass, momentum
and internal energy are in turn:
\begin{equation}
  \label{eq:div mass flux}
  \del \cdot(\avg{\rho}\vec{v}) = 0 ,
\end{equation}
\begin{equation}
  \label{eq:momentum}
  \begin{array}{c}
  \displaystyle \avg{\rho}\left[ \frac{\partial\vec{v}}{\partial t} +
    (\vec{v} \cdot \del)\vec{v} +
    2 \vec{\Omega}_0 \cross \vec{v} \right] 
  =  \\[3mm]
 \displaystyle -\del (\avg{P} + P) + (\avg{\rho} + \rho) \vec{g} -\del \cdot \scrD ,
  \end{array}
\end{equation}
\begin{equation}
  \label{eq:entropy}
  \begin{array}{ll}
  \displaystyle \avg{\rho}\avg{T}\frac{\partial S}{\partial t}
  =
  & \displaystyle \del \cdot \left[ \kappa_r \avg{\rho} c_p \del(\avg{T}+T) + 
                    \kappa_0 \avg{\rho} \avg{T} \del \avg{S} +
		    \kappa \avg{\rho} \avg{T} \del S \right] \\[3mm]
  & \displaystyle - \avg{\rho} \avg{T} \vec{v} \cdot \del(\avg{S}+S) 
  + 2 \avg{\rho}\nu \left[e_{ij}e_{ij} - \frac{1}{3}(\del \cdot \vec{v})^2\right],
  \end{array}
\end{equation}
where $\vec{v} = (v_r, v_\theta, v_\phi)$ is the local velocity in the
rotating frame of constant angular velocity $\vec{\Omega}_0$, 
$\vec{g}$ is the gravitational acceleration, 
$c_p$ is the specific heat at constant pressure, 
$\kappa_r$ is the radiative diffusivity and $\scrD$ is the viscous
stress tensor, given by
\begin{equation}
  \scrD_{ij} = -2 \avg{\rho} \nu \left[e_{ij} 
    - \frac{1}{3}(\del \cdot \vec{v})\delta_{ij} \right],
\end{equation}
where $e_{ij}$ is the strain rate tensor.  Here $\nu$
and $\kappa$ are the diffusivities for vorticity and entropy.  
We assume an ideal gas law
\begin{equation}
  \avg{P} = \scrR \avg{\rho} \avg{T},
\end{equation}
where $\scrR$ is the gas constant, and close this
set of equations using the
linearized relations for the thermodynamic fluctuations
\begin{equation}
  \frac{\rho}{\avg{\rho}} = \frac{P}{\avg{P}} - \frac{T}{\avg{T}}
    =  \frac{P}{\gamma \avg{P}} - \frac{S}{c_p}.
\end{equation}
The mean state variables are evolved with the fluctuations, thus allowing
the convection to modify the entropy gradients which drive it.

ASH is a large-eddy simulation (LES) code, with subgrid-scale (SGS)
treatments for scales of motion which fall below the spatial resolution in our
simulations.  We treat these scales with effective eddy diffusivities,
$\nu$ and $\kappa$, which represent momentum and heat transport by
unresolved motions in the simulations. Here for simplicity $\nu$ and $\kappa$ are taken
as functions of radius alone, and proportional to
$\bar{\rho}^{-1/2}$.  These simulations are similar to
case~AB as reported in \cite{Brun&Toomre_2002} though with a
different SGS functional form (there $\nu, \kappa \propto \bar{\rho}^{-1}$),
and we shall consider faster $\Omega_0$.
Our adopted SGS variation here, as in \cite{Brun_et_al_2004},
\cite{Browning_et_al_2006} and \cite{Ballot_et_al_2007}, yields lower
diffusivities near the top of the layer and thus higher Reynolds numbers.
The thermal diffusion acting on the mean entropy gradient $\kappa_0$
is treated separately and occupies a narrow region in the upper
convection zone.  Its purpose is to transport heat through the outer
surface where radial convective motions vanish.

Our simulations are still separated
by many orders of magnitude from the intensely turbulent conditions
present within the solar convection zone.  They are
likely to capture many aspects of the dynamics of solar convection, and
we are encouraged by the success that similar simulations 
\citep[e.g.,][]{Miesch_et_al_2000,Brun&Toomre_2002,Miesch_et_al_2006,Miesch_et_al_2008}
have had in beginning to match the detailed observational constraints for
differential rotation within the solar convection zone provided by
helioseismology \citep[c.f.\ ][]{Thompson_et_al_2003}.


\begin{deluxetable*}{ccccccccccc}
   \tabletypesize{\footnotesize}
    \tablecolumns{11}
    \tablewidth{0pt}  
    \tablecaption{Parameters for Primary Simulations
    \label{table:sim parameters}}
    \tablehead{\colhead{Case}  &  
     \colhead{ $N_r, N_\theta, N_\phi$} &
     \colhead{Ra} &  
     \colhead{Ta} &  
     \colhead{Re} &  
     \colhead{$\mathrm{Re}'$} &  
     \colhead{Ro} &
     \colhead{Roc} &   
     \colhead{$\nu$} &  
     \colhead{$\kappa$} &  
     \colhead{$\Omega_0/\Omega_\sol$}
   }
   \startdata
                 G1 & $96 \times 256 \times 512$ & 
		      $ 3.22 \times 10^{4} $ & $ 3.14 \times 10^{5} $ &
		      $\phn\phn  84$ & $ \phn 63$ & $  0.92$ & $  0.61$ & $  2.75 $ & 
		      $ 11.0  $ & $ 1$ \\
                 G2 & $96 \times 256 \times 512$ & 
		      $ 1.75\times 10^{5} $ & $ 3.21\times 10^{6} $ & 
		      $\phn 205$ & $ \phn 85$ & $  0.55$ & $  0.45$ & $  1.72 $ &
		      $  6.87  $ & $ 2$ \\
                 G3 & $96 \times 256 \times 512$ & 
		      $ 4.22\times 10^{5} $ & $ 1.22\times 10^{7} $ & 
		      $\phn 326$ & $ 103$ & $  0.41$ & $  0.36$ & $  1.32  $ & 
		      $  5.28  $ & $ 3$ \\
                 G4 & $96 \times 256 \times 512$ & 
		      $ 7.89\times 10^{5} $ & $ 3.18\times 10^{7} $ & 
		      $\phn 433 $ & $ 119$ & $  0.33$ & $  0.31$ & $  1.09  $ & 
		      $  4.36  $ & $ 4$ \\
                 G5 & $96 \times 256 \times 512$ & 
		      $ 1.29\times 10^{6} $ & $ 6.70\times 10^{7} $ & 
		      $\phn 543$ & $ 133$ & $  0.28$ & $  0.27$ & $  0.94  $ & 
		      $ 3.76  $ & $ 5$ \\
                 G7 & $192 \times 512 \times 1024$ & 
		      $ 2.63\times 10^{6} $ & $ 2.06\times 10^{8} $ & 
		      $\phn 763$ & $ 154$ & $  0.22$ & $  0.22$ & $  0.75  $ & 
		      $  3.01  $ & $ 7$ \\
                G10 & $192 \times 512 \times 1024$ &
		      $ 5.58\times 10^{6} $ & $ 6.74\times 10^{8} $ & 
		      $1051$ & $ 188$ & $  0.17$ & $  0.18$ & $  0.59  $ & 
		      $  2.37  $ & $10$ \\[0.25cm]
                G3a & $96  \times 256 \times 512$ &
		      $ 7.83\times 10^{5} $ & $ 2.41\times 10^{7} $ & 
		      $\phn 528$ & $ 158$ & $  0.50$ & $  0.34$ & $  0.94  $ & 
		      $  3.76  $ & $ 3$ \\
                G3b & $192 \times 256 \times 512 \phn$ &
		      $ 2.26\times 10^{6} $ & $ 8.02\times 10^{7} $ & 
		      $1121$ & $ 324$ & $  0.70$ & $  0.32$ & $  0.52  $ & 
		      $  2.06  $ & $ 3$ \\
                G5b & $192 \times 512 \times 1024$ & 
		      $ 4.03\times 10^{6} $ & $ 2.23\times 10^{8} $ & 
		      $1347$ & $ 274$ & $  0.41$ & $  0.26$ & $  0.52  $ & 
		      $  2.06  $ & $ 5$ \\[-0.25cm]
     \enddata
     \tablecomments{All simulations have inner radius 
	$r_\mathrm{bot} = 5.0 \times 10^{10}$cm and outer radius of 
        $r_\mathrm{top} = 6.72 \times 10^{10}$cm, with 
	$L = (r_\mathrm{top}-r_\mathrm{bot}) = 1.72 \times 10^{10}$cm
	the thickness of the spherical shell.
	Evaluated at mid-depth are the
	Rayleigh number $\mathrm{Ra} = (-\partial \rho / \partial S)
	(\mathrm{d}\bar{S}/\mathrm{d}r) g L^4/\rho \nu \kappa$, 
	the Taylor number $\mathrm{Ta} = 4 \Omega_0^2 L^4 / \nu^2$, 
	the rms Reynolds number $\mathrm{Re}  = v_\mathrm{rms} L /\nu$ and
	fluctuating Reynolds number $\mathrm{Re}' = v_\mathrm{rms}' L /\nu$,  
	the Rossby number $\mathrm{Ro} = \omega / 2 \Omega_0$ ,
	and the convective Rossby number 
	$\mathrm{Roc} = (\mathrm{Ra}/\mathrm{Ta} \mathrm{Pr})^{1/2}$.
	Here the fluctuating velocity $v'$ has the differential rotation removed: $v' = v -
	\langle v \rangle$, with angle brackets denoting an average in longitude.
	The Prandtl number $\mathrm{Pr} = \nu / \kappa$ is 0.25 for all simulations.  
	The viscous and thermal diffusivity, $\nu$ and $\kappa$, are
	quoted at mid-depth (in units of $10^{12}~\mathrm{cm}^2\mathrm{s}^{-1}$).
        The rotation rate of each reference frame $\Omega_0$ is in multiples
        of $\Omega_\sol=2.6 \times 10^{-6}~\mathrm{rad}~\mathrm{s}^{-1}$ or $414$ nHz.  The~viscous
	time scale at mid-depth $\tau_\nu = L^2/\nu$ is $1250$~days for case G1 and
	is $3640$~days for case G5.  Additional cases considered other
	rotation rates at $1.25, 1.5, 1.75$ and $6~\Omega_\sol$.     
	}
\end{deluxetable*}


We shall impose
velocity boundary conditions which impose no net torques on the shell.
These are in turn:

\begin{enumerate}
  \item Impenetrable top and bottom: $v_r = 0$,
  \item Stress-free top and bottom:
	\begin{equation}
    (\partial/\partial r)(v_\theta/r) =
    (\partial/\partial r)(v_\phi/r) = 0,
	\end{equation}
  \item Constant entropy gradient at top and bottom: 
    $\partial \bar{S}/\partial r = \mathrm{const}$.
\end{enumerate}

Some of our simulations are initialized by perturbing a quiescent state in solid
body rotation.  The growth of convection leads to velocity
correlations that serve to redistribute
angular momentum within the shell, building a differential rotation
and meridional circulation.  We evolve the simulation for long
periods compared variously to convective overturning times, rotation periods or typical
diffusive times.  Other simulations were
started from these evolved states and then run for long epochs
after all adjustments have been made to the frame rotation rate
$\Omega_0$ and to the viscous and thermal diffusivities. 

\subsection{Studies of Global-Scale Convection in G-type Stars}

\begin{figure}[tb]
  \plotone{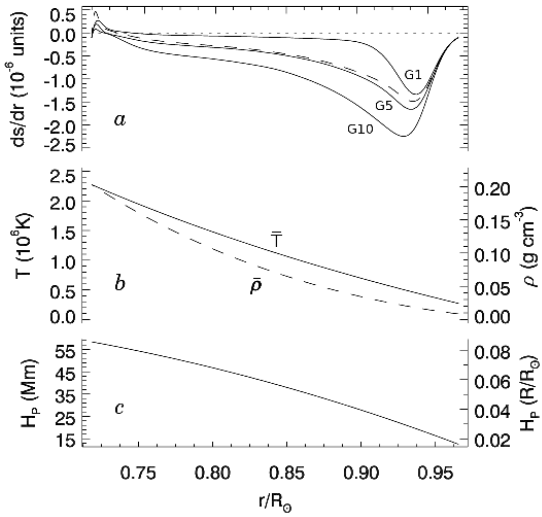}
  \caption{Radial variation of mean stellar structure in the ASH models.
  ($a$) Entropy gradient ($\mathrm{d}\bar{S}/\mathrm{d}r$) for cases~G1,
  G5 and G10 (as labeled).  At higher rotation rates the entropy gradient
  becomes steeper throughout the convection zone, even for our most
  turbulent cases (case~G5b, long dashes).  
  ($b$)  Temperature and density (latter ranging from 0.203 to 0.008
  $\mathrm{g}\thinspace \mathrm{cm}^{-3}$ in the region 
  simulated) for case~G1.  
  ($c$) Pressure scale height $H_P$ (in Mm and fractional solar
  radii), for case~G1, with cases~G2-G10 similar in their mean stratification.
  \label{fig:ash_structure}}
\end{figure}

Our numerical model is a relatively simple description of the solar
convection zone that captures the essential spherical geometry and
global connectivity of that domain.  Solar values are taken for heat
flux, mass and radius and a perfect gas is assumed.  Near the solar
surface the H and He ionization zones, coupled with radiative losses, 
drive intense convective motions
on very small scales which appear at the surface as granulation.
Capturing granulation in a global simulation would require
spherical harmonic degrees of order 4000 and this is currently too
demanding.  We therefore position the top of our domain
slightly below these ionization layers.  In
these simulations we also omit the stably stratified
radiative interior and the shear layer at the base of the convection
zone known as the tachocline.  We focus here on the bulk of the
convection zone, with our computational
domain extending from $0.72R_{\sol}$ to $0.96R_{\sol}$, thus spanning
172~Mm in radius.  The reference
or mean state of our thermodynamic variables is derived from a
one-dimensional solar structure model \citep{Brun_et_al_2002} and is
continuously updated with the spherically symmetric components of the
thermodynamic fluctuations as the simulations proceed.  
These values are illustrated in Figure~\ref{fig:ash_structure} after
convection has readjusted the stratification.

Our studies here explore a variety
of solar-like stars rotating from $1$ to $10~\Omega_\sol$
(cases~G1-G10).  All cases use the same initial stellar structure.  
We seek here to explore the general effects of rotation on stellar convection
rather than the evolution of a particular star, which would
require modifications to the stellar structure as the star aged.  In
surveying the effects of more rapid rotation on global-scale
convection, we sought to achieve reasonably high levels of turbulence
in the resulting flows.  Thus our trajectory through the parameter
space of $\Omega_0, \nu,$ and $\kappa$ attempts to maintain strong
nonlinearity without having increasing the $\Omega_0$ laminarize the convection.
As we increased the rotation rate, we simultaneously decreased the
effective eddy diffusivities $\nu$ and $\kappa$ to maintain the
supercriticality of the simulated turbulent convection.  
We note that the critical Rayleigh number for the onset of
convection scales with rotation as $\mathrm{Ra}_c \propto \Omega_0^{4/3}$ for
Boussinesq convection \citep[e.g.,][]{Chandrasekhar_1961,
Dormy_et_al_2004}.  Lower diffusivities lead to both longer viscous and
thermal diffusion time scales and to flows possessing finer spatial
scales.  Achieving equilibrated states in these systems requires high resolution
simulations carried out over extended periods.  We have taken a middle ground
between attempting to maintain constant supercriticality (which may
require scaling $\nu, \kappa \propto \Omega_0^{-2}$) and keeping
resolution requirements reasonable by scaling our diffusivities as
$\nu, \kappa \propto \Omega_0^{-2/3}$.  However, all of our cases here
are still highly supercritical, noting that the critical Rayleigh number for these
simulations at $1~\Omega_\odot$ is 
$\mathrm{Ra}_c \thinspace \sim \thinspace 1000$
\citep{Gilman&Glatzmaier_1981, Miesch_thesis}.
We have maintained a constant Prandtl number $\mathrm{Pr} = \nu
/ \kappa = 0.25$ in all of our simulations.  The parameters of our
models are detailed in Table~\ref{table:sim parameters}.  Our choice of scalings
for the eddy diffusivities with rotation rate may have some influence
on the nature
of the convective patterns and mean flows we achieve. To assess some of the
sensitivity the choice of our path through parameter space, we
have also sampled a limited range of more turbulent simulations at a
few rotation rates (cases G3a, G3b and G5b).

All of the simulations discussed here are at approximately the same level of
maturity in their evolution.  Case~G1 was the progenitor case at
$1~\Omega_\sol$ and was evolved for some
3000 days after branching away from case~AB from
\cite{Brun&Toomre_2002}, which itself has seen about 10000 days of
total simulated life.  Starting with this case, each subsequent
simulation was spun up from the next fastest case (i.e., G3 was spun up
from G2) and evolved for over 4000 days, or many hundreds of rotation
periods.  At this point, all cases appear to be statistically
stationary in terms of the angular momentum fluxes and the kinetic
energies.  We believe that the differential rotation profiles presented
are effectively stationary, though there are small
fluctuations as determined from
short averages over a few rotation periods.  Certain cases
(including~G5) were evolved for much longer intervals (over 10000 days
and more than 2000 rotation periods) to explore the long-term behavior
of convective patterns in these rapidly rotating systems.  To test
that our results are not unduly subject to hysteresis in the system,
we explored a branch of cases which were successively spun down from
$5~\Omega_\sol$ to $1~\Omega_\sol$.  No significant hysteresis was
found.


\section{Convection in Rapidly Rotating Suns}
\label{sec:convection}

\begin{figure}
  \plotone{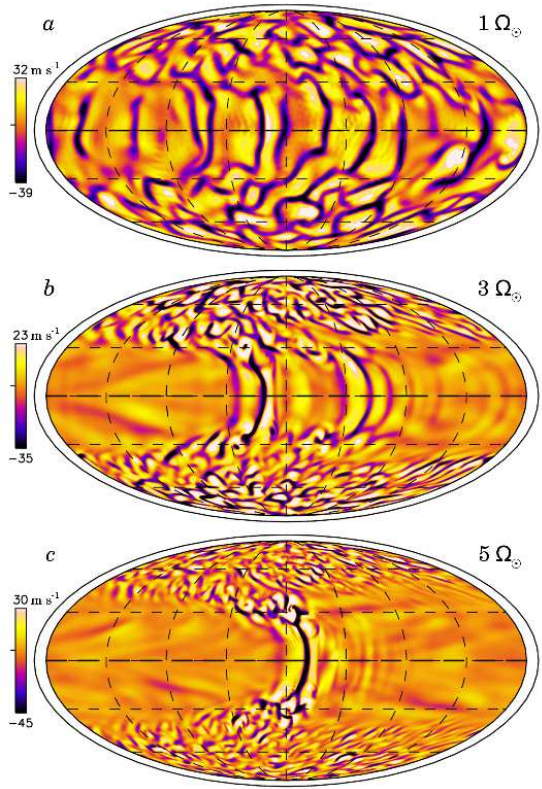}
  \caption{Changes in convective patterns with increasing
  rotation rate for mildly turbulent cases at (\emph{a})
  one, (\emph{b}) three and (\emph{c}) five
  times the solar rate.  Shown as snapshots are radial velocities near
  top of layer in global Mollweide projection, with upflows light and
  downflows dark. Poles are at top and bottom,
  and the equatorial region appears at middle, with equator indicated
  by bold dashed line. Thin dashed lines denote circles of constant
  latitude or longitude, and the thin surrounding line indicates the
  location of the stellar surface at $R_\sol$.  A striking pattern of 
  convection localized into nests near the equator emerges as the
  rotation rate increases. 
  \label{fig:ab2}}
\end{figure}

\subsection{Early Results of Modulated Convection}
In our previous simulations of rapidly rotating suns we found that
strongly localized states of convection emerged with more rapid
rotation \citep{Brown_et_al_2004}.  A selection of these simulations
in Figure~\ref{fig:ab2} present snapshots of
radial velocity near the top of the domain in global Mollweide
projection, showing the entire spherical surface with minimal distortion.
With more rapid rotation, a prominent
longitudinal modulation appears in the patterns of equatorial
convection.  At the higher rotation rates the equatorial convection is
confined to one or two nests, with streaming zonal flow
filling the regions outside these nests of convection.  These nests
can persist for intervals spanning many hundreds of rotation periods,
often with little change.  Two nest states sometimes evolve into
single nest states as one nest overtakes another.  

The simulations shown in Figure~\ref{fig:ab2} are less turbulent than
the cases presented in the rest of this paper, each possessing
Reynolds numbers that are about three-fold smaller near the surface
than in our new simulations. In the former models, a
large portion of the energy transport in the upper convection zone was
carried by unresolved scales of motion.  This parametrized SGS flux,
represented by $\kappa_0$, dominated transport in the upper 30\% of
the convection zone, leading to weaker enthalpy transport and
weaker resolved convection there.  This parametrized flux
acts as volume cooling term that removes flux from the regions where
it is dominant; the dynamics were influenced by the presence
of this cooling layer.  The cases presented in this paper have a 
narrower unresolved flux layer, confined now to the upper 10\% of the
convection zone, and consequently much more vigorous convection is
realized throughout the domain.
In these more turbulent cases, the phenomena of localized
nests of convection is realized at somewhat higher rotation rates. 

%
%

\begin{figure*}[p]
  \plotone{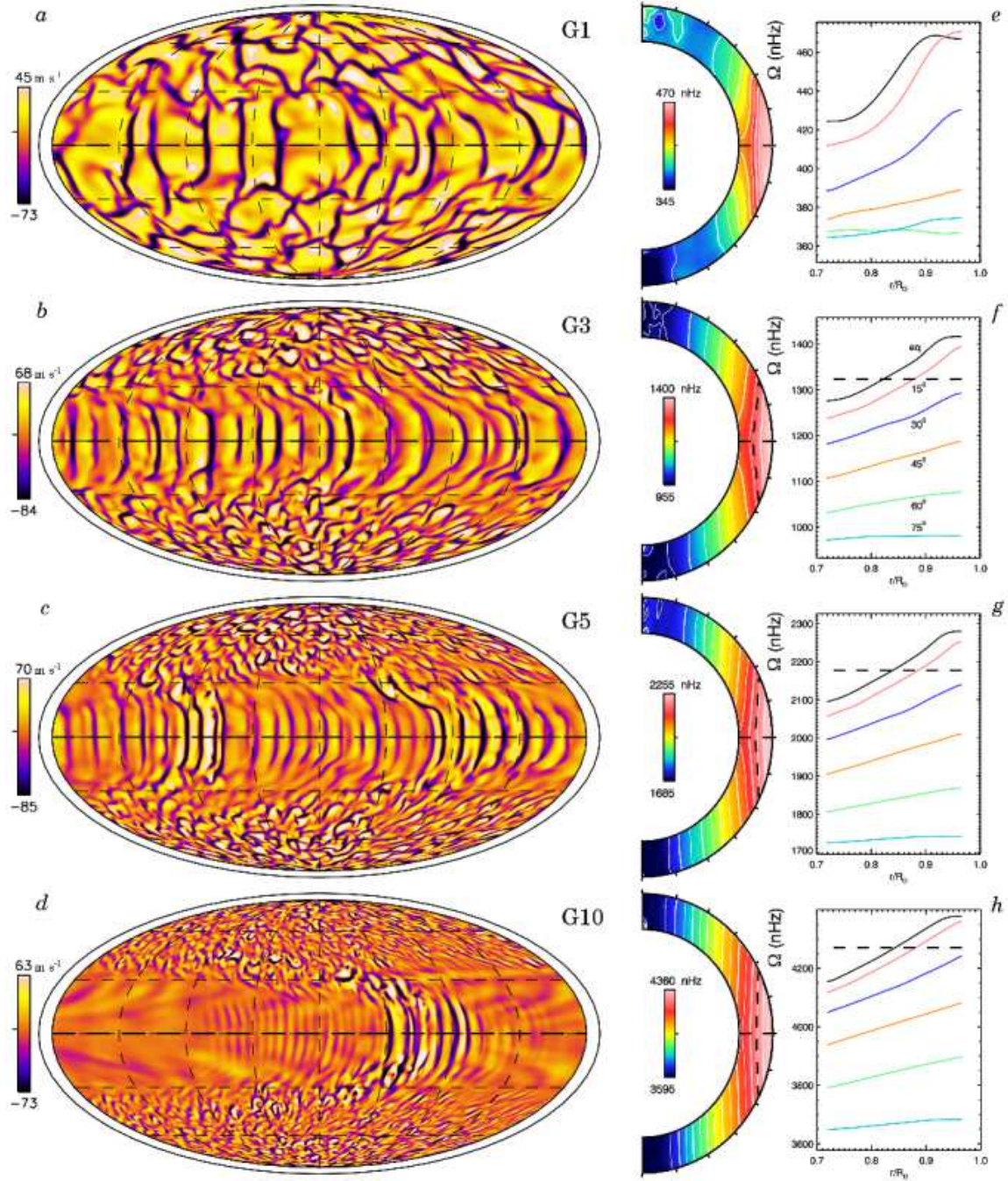}
  \caption{Radial velocity patterns in Mollweide projection at
  $0.95R_\sol$ (\emph{left}) and differential rotation profiles
  (\emph{middle, right}) with increasing rotation rate in (\emph{a,e})
  for~case~G1, (\emph{b,f}) for~G3, (\emph{c,g}) for~G5, and
  (\emph{d,h}) for~G10.  At higher rotation rates the horizontal scale
  of convective cells shrinks at all latitudes and cells are more
  strongly aligned with the rotation axis.  A striking pattern of
  modulated convection emerges at low latitudes with faster rotation,
  consisting of spatially modulated or patchy convection.  These
  active nests of convection are propagating structures which persist
  for long periods of time.  At \emph{middle} are profiles of mean angular
  velocity $\Omega$ with radius and latitude.  These differential
  rotation profiles all involve fast equators (prograde relative to
  the frame rate $\Omega_0$, indicated by tickmark on scale) and a monotonic decrease of $\Omega$ as
  the poles are approached.  At \emph{right} are radial cuts of the angular
  velocity at selected latitudes, as labeled.  The dark dashed contour
  denotes the constant propagation rate of the nests where
  discernible. 
  \label{fig:ab2_turf}}
\end{figure*}


\subsection{Convective Patterns and Evolution with Rotation}
The variation of convective patterns with increasing rotation rate
$\Omega_0$ in our more rapidly rotating suns
is illustrated in Figure~\ref{fig:ab2_turf}.  Snapshots of the radial velocity near
the top of the domain ($0.95 R_\sol$) are shown in Mollweide
projection for four cases: G1, G3, G5 and G10.
The convection patterns are complex and time dependent, with asymmetries
between the upflows and downflows
owing to the density stratification:  narrow, fast downflow lanes are
surrounded by broad, relatively weak upflows.

There is a clear difference in both the scale and structure of
convection at high and low latitudes.  In the equatorial regions
(roughly $\pm30^\circ$ in latitude), the downflows organize into large
structures (loosely called banana cells) aligned
with the rotation axis, thus extending in the north-south direction.  At high rotation rates this tendency
for alignment becomes pronounced, largely in the spirit of the
Taylor-Proudman theorem, and the downflow network exhibits
little of the east-west branching visible in case~G1.
These downflow lanes propagate in a prograde sense relative
to the bulk rotation rate and do so more rapidly than the differential
rotation which they themselves establish.
The nests of convection,
when they appear at the higher rotation rates, propagate at an
intermediate rate as denoted by the heavy dashed contours in
Figure~\ref{fig:ab2_turf}\emph{f-h}.  We defer discussion of the nature of
these nests of convection until later.
Individual convective cells persist for about 10 to 30 days.   

In the higher latitude regions, the convection cells are more isotropic and the
downflow network organizes on smaller scales.
Convection in these regions is vigorous and complex, with upflows
and their downflow networks in a constant dance.
The convective cells have a cusped appearance, with downflows leading
upflows as both propagate in a retrograde fashion (most apparent in
Figs~\ref{fig:ab2_turf}$b,c$). 
Strong vortical plumes form in the interstices of the downflow network
at both high and low latitudes.  In the polar regions above the
middle of the convection zone, the sense of
vorticity in these downflow plumes is generally
cyclonic: counterclockwise in the northern
hemisphere and clockwise in the southern.  As they descend through the
mid-layer their vorticity changes and they become largely
anti-cyclonic.  In contrast, the polar upflows are anti-cyclonic
at all depths.   

The latitudinal variation of convection patterns can be in part understood by
considering a cylinder tangent to the base of the convection zone and aligned with
the rotation axis.  Within our geometry, this tangent
cylinder intersects the outer boundary at about $\pm 42^\circ$ of
latitude.  It is well known that in a rotating convective shell the
flow dynamics are different inside and outside of the tangent
cylinder, owing to differences in the connectivity of the flows, the
influence of the Coriolis forces and distance from the rotation axis
\citep[e.g.,][]{Busse_1970}.  These differences become more evident as
the rotation rate, and hence the rotational constraints on the
convection, increases.
With more rapid rotation the longitudinal extent of the convective cells
becomes progressively smaller.
Linear theory, in the Boussinesq approximation, predicts that the
wavenumber of the most unstable mode scales with rotation as
$m \propto \mathrm{Ta}^{1/6} \propto \Omega_0^{1/3}\nu^{-1/3}$ \citep[e.g.,][]{Chandrasekhar_1961,
  Dormy_et_al_2004} for polar and equatorial convection.  This effect
is found in anelastic systems as well \citep{Glatzmaier&Gilman_1981}
and in our simulations is evident at both high and low latitudes.  

Shown at right in Figures~\ref{fig:ab2_turf}\emph{e-h} are the
profiles of differential rotation (as angular velocity $\Omega$)
realized in these simulations.  These $\Omega (r,\theta)$ profiles are
azimuthally and temporally averaged over a period of roughly
200~days.  All of our more rapidly rotating solar-like stars exhibit
solar-like differential rotation profiles, with prograde (fast)
equators and retrograde (slow) poles.  Contours of constant angular
velocity are aligned nearly on cylinders, influenced by the
Taylor-Proudman theorem, though recent simulations of solar convection
suggest that this is sensitive to the treatment of the bottom
thermal boundary condition \citep{Miesch_et_al_2006}.  
As first shown in mean-field models by \cite{Rempel_2005} and then in
models of global-scale convection by \cite{Miesch_et_al_2006},
introducing a weak latitudinal gradient of entropy at
the base of the convection zone, consistent with a thermal wind balance in a
tachocline of shear, can serve to rotate the $\Omega$ contours toward the more
radial alignment deduced from helioseismology without significantly
changing either the overall $\Omega$ contrast with latitude or the
convective patterns.  We expect similar behavior here, as briefly
explored by \cite{Ballot_et_al_2007}, but have not
explored this issue in detail at the higher rotation rates.
More rapidly rotating suns may very well also possess tachoclines,
but at this stage there is no observational evidence of this.
Thus we have simplified these simulations by imposing a constant entropy
at the bottom boundary.
Our contours of $\Omega$ in Figure~\ref{fig:ab2_turf} 
show some differences between the northern and southern
hemispheres, particularly at higher latitudes, and these differences
decrease with more rapid rotation.  The patterns of convection are not
simply symmetric about the equator, and thus the accompanying mean
zonal flows can be expected to show some variations between the
two hemispheres.  Also shown are radial cuts of $\Omega$ at six fixed
latitudes that make evident the angular velocity contrasts with radius
and latitude achieved in these simulations.  The absolute contrast in
latitude and radius clearly grows with rotation rate, and will be
discussed in \S\ref{sec:global scale flows}.

A most striking result of our simulations is the emergence of
persistent, spatially modulated convection in the equatorial regions
at high rotation rates.  At these low latitudes, convection becomes
modulated in longitude and forms distinct active nests where the convective
vigor is enhanced compared to the regions outside.  The amplitude
of the convective motions and enthalpy transport is larger within
these nests, and indeed at the highest rotation rates,
convection in the equatorial band is confined entirely to the nests.
These nests of convection propagate at a velocity distinct from either
the zonal flow of differential rotation or that of the individual
cells of convection and persist for very long periods of time (more
than 5000~days in case~G5). The nature of these active nests of
spatially localized convection will be explored in detail in
\S\ref{sec:patches}.

Weak modulation in longitude is already evident at low rotation rates.
When long time series are considered we have positively identified
nests of convection in all simulations rotating at 
$\Omega_0 \gtrsim 3~\Omega_\sol$.  As the rotation rate increases, the
modulation level gradually increases; at the highest rotation rates
($\gtrsim 7~\Omega_\sol$) the equatorial convection is almost solely
confined to the nests.  The convection realized in case~G10
(Fig.~\ref{fig:ab2_turf}$d$) is marked by this extreme modulation,
with strong upflows and downflows inside the nest and very little
convection in the surrounding regions.  These most rapidly rotating
cases maintain a strong differential rotation profile, even though the
equatorial convection occupies only a narrow interval in longitude.
The regions outside the nest are filled with fast streaming zonal
flows consistent with the differential rotation.

\begin{figure}  
  \plotone{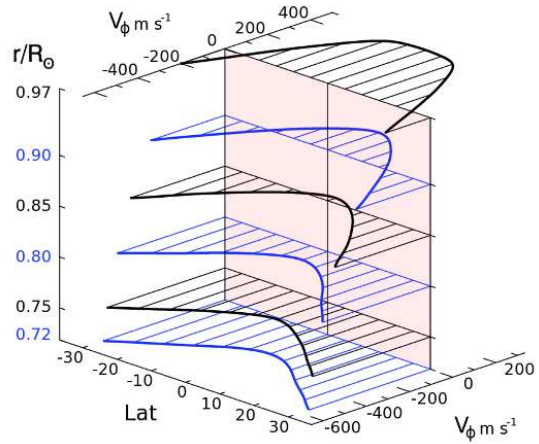}
  \caption{Variation of mean zonal velocity $\langle v_\phi \rangle$
  with latitude for case~G5 sampled at six radial cuts as labeled and
  shown here relative to the uniform propagation rate of the nests of
  convection.  The nests experience a strong prograde zonal flow
  (positive) near the top of layer and a prominent retrograde flow
  within the lower half.
  \label{fig:G5_patch_shear}}
\end{figure}

\subsection{Radial Connectivity of Convection}
The nests of enhanced convection span the
convection zone and propagate everywhere at a constant prograde angular velocity
relative to the bulk rotation rate of the star.  A contour
corresponding to this characteristic propagation rate is overplotted
on the differential rotation profiles in Figures~\ref{fig:ab2_turf}\emph{f-h}
for cases~G3, G5 and~G10.  As is
evident from these profiles, the angular velocity associated with the differential rotation
exceeds that of the propagation rate of the nests near the surface and
is slower than that near the base of the convection zone.  The nests of convection
therefore live within an environment of substantial zonal shear with radius, as is
quantified for case~G5 in Figure~\ref{fig:G5_patch_shear}.  Here the
shearing zonal velocity of differential rotation is plotted in latitude at six radial
depths.  At all depths there is substantial zonal flow through the nests of
convection.  

\begin{figure}
  \plotone{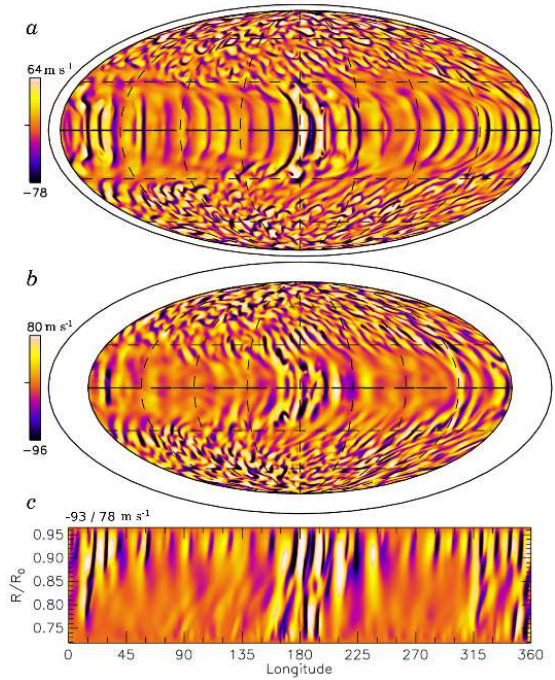}
  \caption{%
    Connectivity of radial velocity with depth in case~G5, shown at
    same instant in Mollweide view
    ($a$) near top of domain ($0.95R_\sol$), 
    ($b$) at mid depth ($0.85R_\sol$), 
    and in ($c$) for an equatorial cut in longitude over full depth range.
    Strong plumes span the convection zone in the equatorial regions
    only within the nest of enhanced convection. The weaker cellular
    flows outside the nests are confined by shear to the upper reaches.
  \label{fig:G5_connectivity}}
\end{figure}

In other studies of solar convection, we have found that strong
downflow lanes extend throughout the entire depth of the domain
\citep{Miesch_et_al_2000,Brun&Toomre_2002,Miesch_et_al_2008}.  In our
more rapidly rotating stars, this connectivity with depth changes
markedly, as is illustrated for case~G5 in
Figure~\ref{fig:G5_connectivity} showing radial velocities throughout
the convection zone.  In these rapidly rotating suns, the strong
variation of mean zonal flow with radius in the equatorial regions
prevents all but the strongest downflows from spanning the convection
zone.  Within the nest of enhanced convection the plumes are able to
traverse the convection zone.  Yet in the quieter regions outside, the
weaker downflow plumes are truncated by shear before reaching the
middle of the convection zone.  It is evident
(Fig~\ref{fig:G5_connectivity}$c$) that the amplitude of
convective motions is pronounced at all depths within the nest of active
convection.

The downflowing plumes are influenced by the strong radial shear and
some break into multiple cells with radius even before the full blown
nests of localized convection emerge, as is evident already in our
simulation rotating at twice the solar rate (case~G2).
When the downflow networks only span a portion of the convection zone
and experience a limited range of the full density stratification, the
importance of compressible effects decreases.  This has important
consequences for the energetics of the simulations, particularly the
radial kinetic energy flux, as will be addressed in
\S\ref{sec:energies}. In contrast, the downflowing plumes in the polar
regions experience much less shear from either the differential
rotation or the relatively weak meridional circulations and continue
to span the entire convection zone depth.

\begin{figure*}
  \plotone{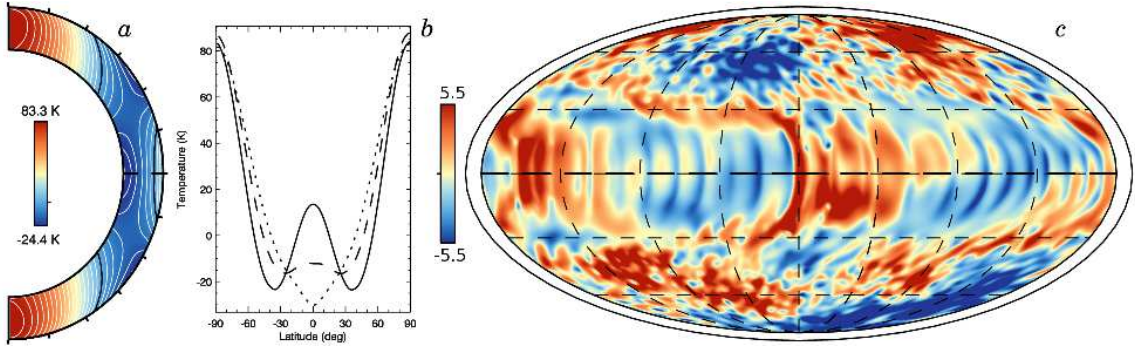}
  \caption{Temperature structures within case~G5.  Mean latitudinal
  variations in temperature are shown relative to their spherical
  average~$\bar{T}$ in ($a$) as contours with radius and latitude and ($b$) as
  cuts at fixed radii at the top (\emph{solid}, $0.96R_\sol$), middle
  (\emph{dashed}, $0.84R_\sol$) and bottom (\emph{dotted}, $0.72R_\sol$)
  of the domain.  ($c$) Temperature fluctuations in a snapshot near top of
  domain ($0.95R_\sol$) relative to the mean structure in ($a$).
  \label{fig:G5_thermal_structure}}
\end{figure*}

\subsection{Thermal Structuring}

In these rapidly rotating suns, the turbulent alignment of convection
with the rotation axis leads to a net latitudinal transport of enthalpy, 
yielding a prominent latitudinal gradient of
temperature.  The resulting thermal structuring in case~G5 is shown in
Figure~\ref{fig:G5_thermal_structure}, presenting both the mean
temperature profile and representative temperature fluctuations in a snapshot
near the surface.  In the latter, individual convection cells are
associated with small fluctuations with amplitudes of a
few K.  Downflows are generally cool while upflows are
relatively warmer.  The enhanced enthalpy transport within the active
nests of convection appears as positive temperature fluctuations
in the equatorial region.    

Evident at high latitudes (Fig.~\ref{fig:G5_thermal_structure}$c$) are
broad spatial structures (in addition to small-scale convection) which
appear in the temperature 
fluctuations and are not readily visible in the maps of radial
velocity (see Fig.~\ref{fig:G5_connectivity} at same instant).
These structures are long lived and appear to be a separate phenomena
from the nests of convection.  The polar patterns propagate in a
retrograde sense more rapidly than the differential rotation in which
they are embedded, and though streaming wakes from the active nests
print weakly into the polar regions, the polar patterns and nests
appear to be distinct phenomena.
The large-scale polar patterns are not evident in the slowly rotating
cases (G1 and G2); in the most rapidly rotating cases this
modulation attains a more complicated form than the two-lobed
structure shown here.

The zonally-averaged thermal structure
(Fig.~\ref{fig:G5_thermal_structure}$a,b$) is quite smooth and is characterized by
warm poles and a cool equator, with yet cooler mid-latitudes.  
In contrast, the mean entropy increases monotonically from
equator to pole, due to effects of pressure.  All of the more rapidly rotating cases
have similar latitudinal thermal profiles, though the temperature
difference between equator and pole increases with more rapid
rotation, as will be discussed further in \S\ref{sec:global scale flows}.
In case~G5, the latitudinal pole to equator temperature contrast is approximately
$100~\mathrm{K}$ throughout the convection zone.  These latitudinal variations remain small
at all rotation rates in comparison to the spherically symmetric
background $\bar{T}$, which ranges from $2.7\times 10^5~\mathrm{K}$ near the surface to
$2.2\times 10^6~\mathrm{K}$ near the bottom of the convection zone (as
shown in Fig.~\ref{fig:ash_structure}).

\section{Building a Strong Differential Rotation}
\label{sec:global scale flows}

\subsection{Thermal Wind Balance}
\label{sec:thermal wind}

Rapidly rotating systems are constrained by the Taylor-Proudman theorem
to have minimal variations in flow dynamics along the direction of the
rotation axis.  In stratified flows, gradients in density and pressure
contribute to baroclinic terms in the vorticity equations
\citep{Pedlosky_1987, Zahn_1992} which maintain flows that can break the
Taylor-Proudman constraint.  In our rapidly rotating suns, convective
plumes tilt towards the rotation axis as rotation effects
increase.  This results in latitudinal as well as radial transport of
enthalpy and builds a latitudinal gradient of temperature and entropy.
Such gradients arise naturally in a rotating convective system even
with uniform thermal boundary conditions.
For a nearly adiabatic stratified, rotating, non-magnetized fluid it
can be shown that in the limit of small Rossby number and negligible
viscous effects the zonal component of the vorticity equations reduces
to the well known thermal wind balance
\citep[e.g.,][]{Brun&Toomre_2002, Miesch_et_al_2006}: 
\begin{equation}
  \frac{\partial \widehat{v_\phi}}{\partial z} = \frac{g}{2 C_P r
  \Omega_0}\frac{\partial \widehat{S}}{\partial \theta} ,
  \label{eqn:TW balance}
\end{equation}
where $z$ is directed along the rotation axis and a hat denotes an
average in longitude and time.  We have further
assumed that the turbulent pressure is negligible.

\begin{figure}
  \plotone{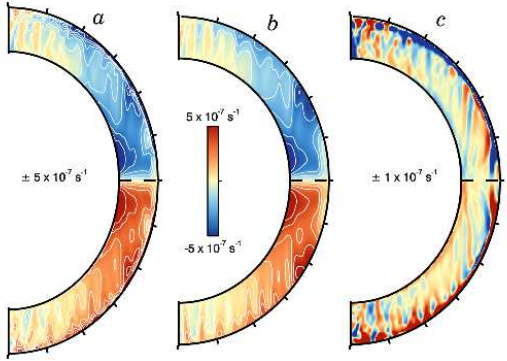}
  \caption{Thermal wind balance achieved in case~G5.  
  ($a$) Gradients of $\widehat{v_\phi}$ along the rotation axis, $\partial \widehat{v_\phi}/\partial z$,
  (\emph{b}) the scaled latitudinal entropy gradient from the
  right-hand side of eq.~(\ref{eqn:TW balance}), and (\emph{c}) their
  difference, with contours in the latter rescaled
  to show the departures near the boundaries.  The bulk of the
  convection zone is in thermal wind balance, but substantial
  departures arise near the top and bottom of the domain where
  Reynolds stresses dominate.
  \label{fig:TW balance}}
\end{figure}

From equation~(\ref{eqn:TW balance}) it is clear that departures from
rotation constant on cylinders (as observed in the solar interior by
helioseismology) can be maintained by a latitudinal gradient of
entropy.  The left and right hand sides of equation~(\ref{eqn:TW
balance}) are shown for case~G5 in Figure~\ref{fig:TW balance}.  In
the bulk of the convection zone, the differential rotation profiles
realized in these more rapidly rotating suns are substantially in
thermal wind balance.  Significant departures arise near the inner and
outer boundaries (Fig.~\ref{fig:TW balance}$c$) 
where Reynolds stresses and boundary conditions play
a dominant role, as was found in earlier simulations of solar
convection \citep{Brun&Toomre_2002}.

\begin{figure}
  \plotone{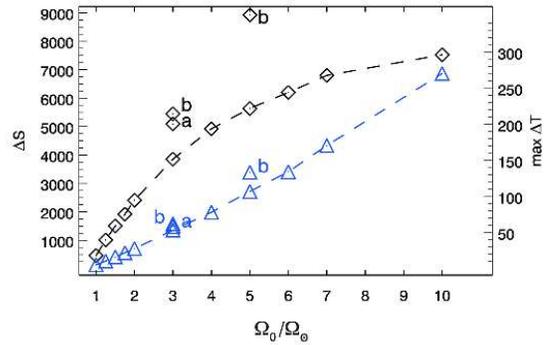}
  \caption{Increase with rotation rate in the latitudinal contrast of
  entropy, $\Delta \widehat{S}$  (plotted as diamonds) between equator and
  high latitudes at $0.96R_\sol$.  The more turbulent cases (G3a, G3b and G5b as labeled)
  have larger entropy contrasts, in keeping with their generally
  stronger differential rotation.  Blue triangles indicate the
  maximum temperature contrast in latitude at the upper boundary in each simulation.  
  \label{fig:thermal_contrast_with_omega}}
\end{figure}

Another striking property of the thermal wind balance is that
increasing $\Omega_0$ leads to more cylindrical profiles of $\widehat{v_\phi}$
unless $\partial \widehat{S}/\partial \theta$ also adjusts with the rotation
rate.  In our more rapidly rotating suns we find that the latitudinal
gradients of temperature and entropy increase with more rapid
rotation.  The growth of $\Delta \widehat{S}$ (difference between the
surface value of $\widehat{S}$ at say $60^\circ$ and the equator)
with increasing rotation rate $\Omega_0$ is shown in
Figure~\ref{fig:thermal_contrast_with_omega}.  The latitudinal
structure of entropy is always monotonic in these simulations, with
lower entropy at the equator and higher entropy at the poles.
Convection in these more rapidly rotating systems establishes stronger
latitudinal gradients, but not enough in these simulations to maintain
the $\Omega$ profiles unchanged.

Accompanying the growth of $\Delta \widehat{S}$ is a growth in the latitudinal
temperature contrast, as shown by the maximum temperature contrast in
latitude near the stellar surface in
Figure~\ref{fig:thermal_contrast_with_omega}.  Typically, the maximal
contrast occurs between the poles and latitudes of $\pm40^\circ$, as
seen in Figure~\ref{fig:G5_thermal_structure} for case~G5 with a
contrast of about $100~\mathrm{K}$.  In the rapidly rotating
simulations, the primary flux balance in latitude is between thermal
eddy diffusion $\kappa \bar{\rho} \bar{T} \langle \partial S/\partial
\theta \rangle$ and convective enthalpy transport $C_p \bar{\rho}
\langle v_\theta' T' \rangle$.
Here convective transport moves warm
material to the poles as the downflows align more strongly with the
rotation axis while eddy diffusion works to erode the gradient.  
The meridional circulations appear to play a
relatively minor role in maintaining the overall latitudinal entropy
contrast.

\subsection{Angular Momentum Redistribution}

In these simulations of stellar convection, complex couplings between
rotation and convection build the profiles of differential rotation and
meridional circulation.
With stress-free boundary conditions at the top and bottom of the
shell there are no net torques and thus the total angular momentum is conserved.
Couplings between rotation and convection lead to a global-scale
redistribution of angular momentum, resulting in the sustained flows of both
differential rotation and meridional circulation.  To assess the
transport of angular momentum in these systems we follow the approach
of \cite{Miesch_et_al_2008}, examining the average radial and latitudinal
angular momentum transport as detailed in their equations~(10)-(12)
\citep[see also][]{Brun&Toomre_2002,Miesch_2005}. 
The angular momentum fluxes from Reynolds stresses (RS), meridional
circulations (MS) and viscous diffusion (VD) are
\begin{eqnarray}
  \vec{F}^{\mathrm{RS}} = \bar{\rho} r \sin{\theta} 
  \left(\langle v_r' v_\phi' \rangle \vec{\hat{r}} +
        \langle v_\theta' v_\phi' \rangle \vec{\hat{\theta}} \right),
	\label{eq:F_RS}\\
  \vec{F}^{\mathrm{MC}} = \bar{\rho} {\cal L} 
  \left(\langle v_r \rangle \vec{\hat{r}} +
        \langle v_\theta \rangle \vec{\hat{\theta}} \right),
	\label{eq:F_MC}\\
  \vec{F}^{\mathrm{VD}} = -\bar{\rho} \nu r^2 \sin^2{\theta} \nabla \Omega,
  \label{eq:F_VD}
\end{eqnarray}
where 
\begin{equation}
{\cal L} = r \sin\theta \left(\Omega_0 r \sin\theta + \left<v_\phi\right>\right)
\end{equation}
is the specific angular momentum.

\begin{figure}
  \plotone{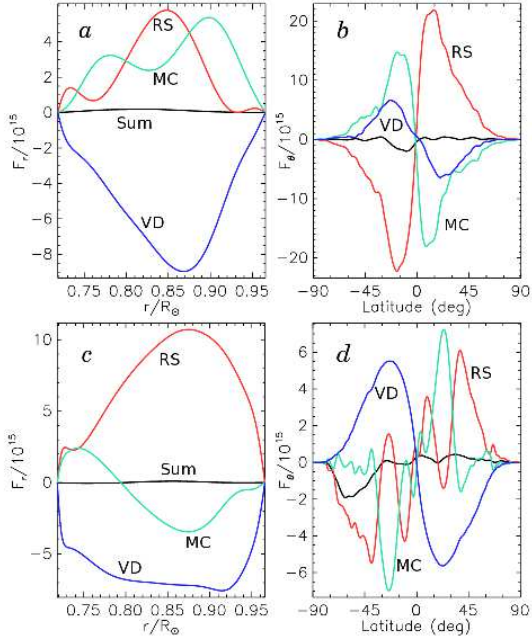}
  \caption{Time average of the integrated radial ($F_r$) and
  latitudinal ($F_\theta$)
  angular momentum flux for case~G1 ($a,b$) and case~G5 ($c,d$).
  Shown are the fluxes from Reynolds stresses (RS), meridional
  circulations (MC), viscous diffusion (VD) and their total.
  Transport by viscous diffusion remains comparable in all cases,
  while the transport by Reynolds stresses and meridional circulations
  changes markedly with more rapid rotation.
  \label{fig:amom_balance}}
\end{figure}

The total radial and latitudinal fluxes of angular momentum are shown
for case~G1 and G5 in Figure~\ref{fig:amom_balance}.  Here we have
integrated in co-latitude and radius respectively to deduce the net
fluxes through shells at various radii and through cones at various
latitudes \citep[c.f.,][]{Miesch_2005}.  The three major contributions
arise from Reynolds stresses,  
meridional circulations and viscous terms.  Velocity correlations
lead to net angular momentum transport by Reynolds stresses as
convective structures develop organized tilts and align partially with
the axis of rotation
\citep[e.g.,][]{Brummell_et_al_1998,Brun&Toomre_2002, Miesch_et_al_2008}.  
This alignment is particularly prominent in the fast downflow lanes,
and becomes stronger as rotation increases.

Turning first to our solar case (G1, Fig.~\ref{fig:amom_balance}$a,b$),
we see that in radius the meridional circulations and Reynolds
stresses play similar and nearly equal roles in transporting angular
momentum outward.  The viscous flux meanwhile is negative and
transports angular momentum inward, in keeping with the positive
radial gradient of the differential rotation profile
(eq.~\ref{eq:F_VD}), and the total flux in radius is nearly zero.
The transport in latitude is somewhat different.  Here
meridional circulations combine with viscous fluxes to transport
angular momentum away from the equator and toward the poles
(i.e., positive in the southern hemisphere and negative in the northern
hemisphere). This tendency is opposed by Reynolds stresses, which
continuously accelerate the equatorial regions and dynamically
maintain the angular velocity contrast $\Delta \Omega$ in latitude.

The transport of angular momentum in our more rapidly rotating cases
are all similar in form and are well represented by case~G5
(Fig.~\ref{fig:amom_balance}$c,d$). 
In these more rapidly rotating stars, the radial balance is dominantly
between the Reynolds stresses transporting angular momentum outward and
viscous terms transporting it inward.  The viscous transport is
similar in magnitude to that of case~G1, though the radial boundary
layers are now much narrower.  The transport by Reynolds stresses is nearly twice
as large, and this likely arises from the strong alignment of
convective structures in both polar and equatorial regions. 
In these stars the weaker meridional circulations become relatively
minor and disorganized players in the radial flux balance, moving
angular momentum outward in some regions of the shell and inward in others.
This opposing behavior between the upper and lower convection zone
arises from the meridional circulations breaking into multiple cells
in radius. 

The balances achieved in the latitudinal transport in case~G5
(Fig.~\ref{fig:amom_balance}$d$) are more complex.  As
the rotation rate has increased, the total viscous transport has remained nearly
constant, with the significantly stronger gradients of angular velocity in
the differential rotation profiles
offset by the lower turbulent diffusivities dictated by our path
through parameter space.  That these two opposing actions should
conspire to produce a nearly constant profile of viscous angular
momentum transport is striking and not intuitive.  This is
particularly apparent when we examine the two other terms in the flux balance. 
The meridional circulations have reversed their role from our solar-like case~G1 and
now work with the Reynolds stresses to accelerate the equator and
spin down the polar regions.  The reduced contribution of the meridional
circulations to the total balance arises as the flows become both
weaker and multi-celled in radius and latitude.  
The smaller transport by Reynolds stresses appears to result from the
destruction by radial shear of some of the downflow plumes.

\begin{figure}[tb]
  \plotone{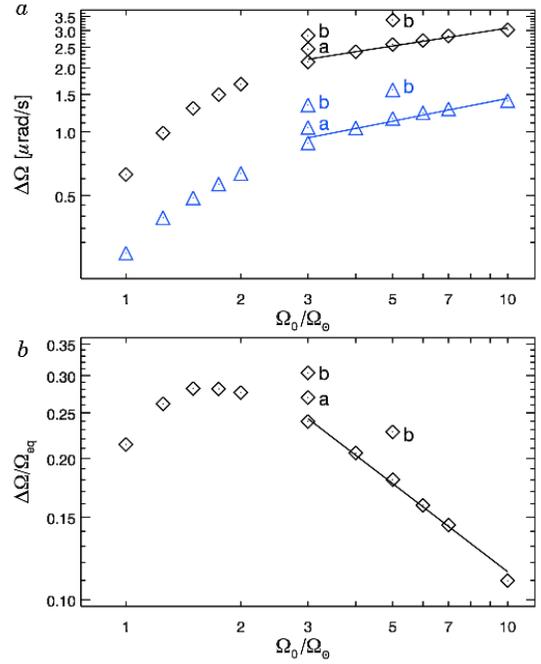}
  \caption{$(a)$ Angular velocity contrast $\Delta \Omega$ in latitude
  between equator and $60^\circ$ (diamonds)
  and in radius across the shell at the equator (blue triangles). 
  The more rapidly rotating cases appear to follow a power law, which
  for the latitudinal contrast is $m=0.3$ and for the radial contrast
  is $m=0.4$ (as in eq.~\ref{eq:absolute_contrast}).
  $(b)$ Relative latitudinal angular velocity contrast $\Delta
  \Omega/\Omega_{\textrm{eq}}$, with the shown power law having $n=-0.6$
  (as in eq.~\ref{eq:relative_contrast}). 
  The scaling may vary with the path in parameter
  space, as suggested by cases~G3a, G3b~and~G5b.
  \label{fig:ratios_of_omega}}
\end{figure}

\subsection{Differential Rotation and Scaling with Rotation}
\label{sec:DR}
In analyzing our simulation results, it is the differential rotation
established by the convection that may yield the most direct contact
with observations.
Stellar observations across the HR diagram indicate that differential
rotation is a common feature in many stars, particularly stars of
spectral class F and later.  In the sun, differential rotation has
been measured throughout the bulk of the convection zone
\citep[as reviewed by][]{Thompson_et_al_2003}, 
but at present for more distant stars only
the surface differential rotation can be inferred.
A variety of observational techniques have been
employed, ranging from photometric variability studies
\citep[e.g.,][]{Donahue_et_al_1996,Walker_et_al_2007}, Doppler imaging
techniques \citep[e.g.,][]{Donati_et_al_2003} and Fourier transform methods
\citep[e.g.,][]{Reiners&Schmitt_2003}.  Typically, these observations seek to measure
the amount of angular velocity contrast at the stellar surface, 
denoted as $\Delta\Omega_*$, though what is being measured may be
somewhat uncertain.  Variations in
$\Delta\Omega_*$ have been found with both rotation rate and spectral
type, however these quantities are correlated and in observations to
date it is difficult to disentangle their possible separate effects 
\citep{Reiners_2006}. 

Different techniques measure fundamentally different tracers of
surface differential rotation, either following variability of
Ca emission (photometric), darkening from inferred starspot presence
(photometric and Doppler imaging) or rotational broadening of absorption lines of
unspotted stars (Fourier transform methods).  Each technique also is most applicable
in only a limited region of stellar parameter space.  As such,
overlapping surveys are in short supply.  Generally,
most observations indicate that the relative
shear, $\Delta\Omega_*/\Omega_*$, depends on the stellar rotation rate
$\Omega_*$ as a power law, though different surveys find different
scalings for the differential rotation, expressed as
\begin{equation}
  \frac{\Delta\Omega_*}{\Omega_*} \propto \Omega_*^n
  \label{eq:relative_contrast}
\end{equation}
(e.g., $n=-0.3\pm0.1$ in \citealt{Donahue_et_al_1996} and 
$n=-0.34\pm 0.26$ in \citealt{Reiners&Schmitt_2003}, but 
$n=-0.85\pm0.10$ in \citealt{Barnes_et_al_2005}). 
Whereas some global models of convection in more rapidly rotating stars have 
been conducted \citep[e.g.,][]{Rudiger_et_al_1998, Kuker&Stix_2001,
  Kuker&Rudiger_2005_A&A, Kuker&Rudiger_2005_AN},
these have been largely carried out in 2-D under the simplifying assumptions
of mean-field theory.

The amount of latitudinal shear observed at the surface $\Delta
\Omega$ is an important quantity both for interpreting stellar
observations and for many 
dynamo theories.  Here we define $\Delta \Omega$ more specifically as the
difference in angular velocity between the equator and say at $60^\circ$
latitude, namely
\begin{equation}
  \Delta \Omega = \Omega_\mathrm{eq} - \Omega_{60} \propto \Omega_0^m.
  \label{eq:absolute_contrast}
\end{equation}
Going to higher latitudes yields comparable behavior.
As shown in Figure~\ref{fig:ratios_of_omega}, we find that
$\Delta \Omega$ increases with rotation rate in our simulations,
with $m=0.3$ in the most rapidly rotating simulations.  The radial shear also
increases with more rapid rotation, and at the equator the difference
between the mean angular velocity at top and bottom of the shell
scales as $m=0.4$ for the rapid rotators.
Because $m<1$, the relative shear $\Delta \Omega / \Omega_\mathrm{eq}$ 
decreases with rotation rate $\Omega_0$ for the rapid rotators, in nearly
a power law fashion for the path through parameter space explored
here.  The scaling exponent from Equation~(\ref{eq:relative_contrast})
exhibited by these cases is $n=-0.6$, but this 
may be influenced by our choice in the scaling of diffusivities with
rotation.  We are 
encouraged that our more turbulent cases~G3b and
G5b, with the same diffusivities at different rotation rates,
exhibit similar behavior.  Our choice of low Prandtl number also has an
effect on this scaling \citep[see ][]{Ballot_et_al_2007}.
Additionally, different treatments of  the SGS unresolved 
flux, which has the most effect in the upper 10\% of the convection
zone, can alter the particular scaling law.  The early simulations
presented in \cite{Brown_et_al_2004} and shown in
Figure~\ref{fig:ab2}, which had a much thicker unresolved flux layer,
had a scaling of $n=-0.8$ in the rapid rotation limit.  We have found
that normalizing $\Delta \Omega$ by $\Omega_0$ rather than
$\Omega_\mathrm{eq}$ leads to a systematic offset for $n$ of about -0.05 in
the inferred scaling law.

\begin{figure}
  \plotone{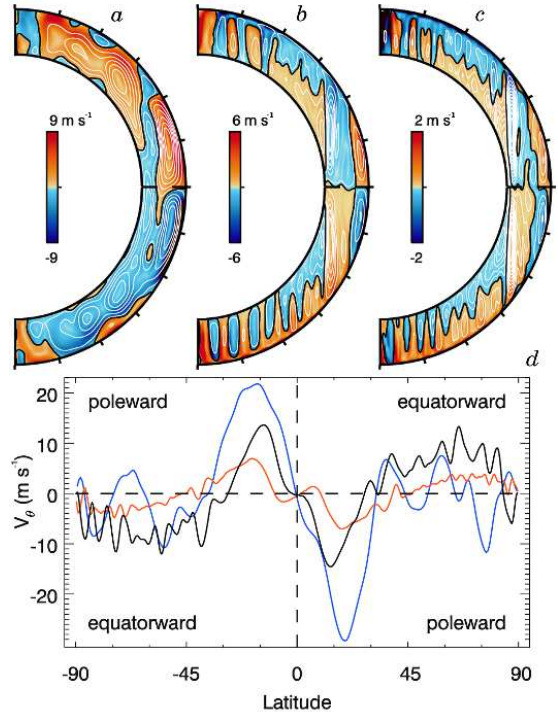}
  \caption{Mean meridional circulations with latitude and radius for
  (\emph{a})~case~G1, (\emph{b})~case~G5 and (\emph{c})~case~G10 with
  streamlines of mass flux $\Psi$ overlaid.  Colors indicate the sense
  (red counter-clockwise, blue clockwise) and magnitude of the
  meridional velocity $\langle \vec{v_m} \rangle = \langle v_r \rangle
  \vec{\hat{r}} + \langle v_\theta \rangle \vec{\hat{\theta}}$.  
  With more rapid rotation the meridional
  circulation cells align strongly with the rotation axis and weaken
  in amplitude.  (\emph{d}) Amplitude of the mean latitudinal component
  $v_\theta$ at the top of the simulation for case~G1~(\emph{blue}),
  G5~(\emph{black}), and G10~(\emph{red}), with regions of poleward
  and equatorward flow denoted.   
  \label{fig:MC}}
\end{figure}

\section{Meridional Circulations and Scaling with Rotation}
\label{sec:MC}
The meridional circulations realized within our simulations are of
significance since they can variously transport heat, angular momentum
and even magnetic fields between the equator and the poles, though the
later are not included in the present simulations.  Our time and
longitude-averaged meridional circulation patterns are shown in
Figure~\ref{fig:MC} for cases~G1, G5 and G10, depicted as streamlines
of mass flux $\Psi$,
\begin{equation}
  r \sin{\theta} \langle \bar{\rho} v_r \rangle = -\frac{1}{r}
  \frac{\partial \Psi}{\partial \theta} 
  \quad \mathrm{and} \quad
  r \sin{\theta} \langle \bar{\rho} v_\theta \rangle = \frac{\partial \Psi}{\partial r}
\end{equation}
and averaged here over a period of at least 150 days.

In our more rapidly rotating cases the meridional circulations have
broken into several cells strongly aligned with the rotation axis
(Fig.~\ref{fig:MC}$b,c$), particularly in the equatorial regions.  Weak
connections between the equatorial and polar regions persist
at the highest rotation rates studied, with organized flows
along the tangent cylinder.  These internal flows weaken with more
rapid rotation.  The meridional circulations are complex and time dependent, with large
fluctuations around the statistically-steady states shown here,
involving variations comparable to or larger than the mean values themselves.  
The circulations are driven by small imbalances between relatively
large forces and their nature is subtle.
The variation of the meridional flows near the surface
($0.96R_\sol$) with rotation rate is shown in Figure~\ref{fig:MC}$d$.  
The amplitude of the
flows decreases substantially with more rapid rotation. Peak
velocities drop from 22~\ms in case~G1 to 14~\ms in G5 and about 7~\ms in G10.  


\begin{figure}
  \plotone{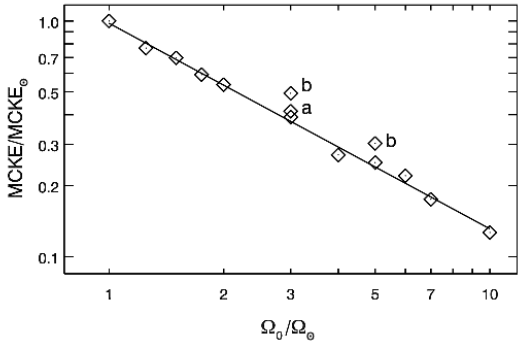}
  \caption{Volume-averaged kinetic energy density of the
  meridional circulations (MCKE) with rotation rate
  $\Omega_0$.  The MCKE is
  normalized by that energy in case~G1 at the solar rate
  ($2.5\times 10^4 \thinspace \mathrm{ergs}~\mathrm{cm}^{-3}$).
  The kinetic energy of these 
  circulations decreases with rotation rate; a power law
  scaling of $\Omega_0^{-0.9}$ is shown for reference.  
   \label{fig:MCKE_vs_omega}}
\end{figure}

The total energy contained in these meridional
circulations decreases quickly with more rapid rotation, as shown in
Figure~\ref{fig:MCKE_vs_omega}.  This drop in energy is independent
of the detailed structure of the convection, showing no change in
behavior at the transition to spatially modulated convection.  In
contrast to the energy contained in convection (CKE) and differential
rotation (DRKE), the energy in the meridional circulations (MCKE) is much
less sensitive to the level of turbulence in any particular
simulation, as indicated by cases~G3, G3a and G3b (detailed in
Table~\ref{table:energy balances}).
The meridional circulations remain important to the global-scale
dynamics as their gradual redistribution of angular momentum
contributes to the large angular velocity gradients in latitude. 
Yet they are inefficient at transporting heat out of the star and at
redistributing thermal material to maintain the latitudinal gradients of
temperature and entropy (which correspond to the thermal-wind component 
of the achieved differential rotation). 

This finding is in striking contrast to the assumptions of many
Babcock-Leighton dynamos, which often take the meridional velocity to
scale as $v_m \propto \Omega$ or $v_m \propto \log{\Omega}$ 
\citep[e.g.,][]{Charbonneau&Saar_2001, Dikpati_et_al_2001}.  
In these models faster meridional circulations lead to shorter dynamo
cycles as surface flux is returned more rapidly to the tachocline.
Currently, the observational data does not appear good enough to
distinguish between the competing models and we will have to await
better measurements of the scaling between cycle period and rotation
rate and possible observations of the meridional circulations
themselves \citep{Rempel_2008}.
Existing 2-D mean-field models of rapidly rotating suns
\citep{Kuker&Stix_2001, Rudiger_et_al_1998, Kuker&Rudiger_2005_A&A}
also predict an increase of meridional circulation velocities with more
rapid rotation.  This is in contrast to their decrease in our simulations.

\section{Energy Balances and Flux Transport}
\label{sec:energies}

\begin{figure}
  \plotone{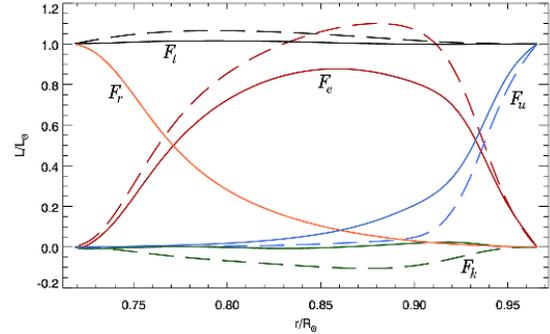}
  \caption{%
  Variation of the energy fluxes with radius for
  cases~G1~(\emph{dashed}) and G5~(\emph{solid}).
  Shown are the radiative
  luminosity ($F_r$), convective enthalpy transport
  ($F_e$), unresolved flux ($F_u$), kinetic energy transport
  ($F_k$), and the total flux through the
  simulation ($F_t$), all normalized by $L_\odot/(4 \pi r^2)$.  
  Case~G1 has a large positive
  convective enthalpy flux throughout the domain, with
  the excess in luminosity largely balanced by an inward flux of kinetic energy.
  At higher rotation rates (as in G5), the kinetic energy flux is nearly zero
  throughout the domain.  In both cases $F_\nu$ is negligible and not shown.
  \label{fig:flux_balance}}
\end{figure}

Convection is responsible for transporting the stellar flux emerging
from the deep interior through the convection zone.  In these
simulations, the total luminosity $L(r)$ and its components are
\begin{equation}
  F_e + F_k + F_r + F_u + F_\nu = \frac{L(r)}{4 \pi r^2} = F_t,
\end{equation}
with
\begin{eqnarray}
  & F_e = c_p \bar{\rho} \overline{v_r T'}, 
  & F_k = \case{1}{2} \bar{\rho} \overline{v_r v^2},\\
  & F_r = -\kappa_r c_p \bar{\rho} \frac{\mathrm{d} \bar{T}}{\mathrm{d} r}, \quad
  & F_u = -\kappa_0 \bar{\rho} \bar{T} \frac{\mathrm{d}\bar{S}}{\mathrm{d}r}, \quad
    F_\nu = -\overline{\vec{v} \cdot \vec{\scrD}} {\big|}_r, \quad ~~
\end{eqnarray}
where $F_e$ is the enthalpy transport by convective motions, $F_k$ is
the kinetic energy flux, $F_r$ is the transport by radiation, $F_u$ is
the unresolved SGS heat flux for parametrized transport by scales of motion below the
resolution of our simulation and $F_\nu$ is the SGS viscous flux.
Figure~\ref{fig:flux_balance} shows the 
flux balance with radius achieved in cases~G1 and
G5, averaged over horizontal surfaces and converted to relative luminosities.   
In the deepest layers the radiative flux becomes significant as the
radiative conductivity steadily increases with depth.  By construction
it suffices to carry all of the imposed flux through the lower
boundary, where the radial velocities and convective flux vanish.
A similar role is played near the top of the convection zone by the
sub-grid scale transport which yields $F_u$.
The main role of $F_u$ is to transport energy outward through
the impenetrable upper boundary where the convective fluxes vanish and
the remaining fluxes are small, thereby avoiding the 
building of strong superadiabatic radial gradients there. 

The functional form of $\kappa_0(r)$ is chosen so that the entire stellar
luminosity will be transported at the surface of the simulation by $F_u$.
A subtlety of this treatment of the SGS
flux lies in $\mathrm{d} \bar{S}/\mathrm{d}r$.  In the more rapidly
rotating simulations, we find that convection is less able to establish
an adiabatic profile throughout the convection zone.  Instead, much of
the convection zone remains slightly superadiabatic 
($\mathrm{d} \bar{S}/\mathrm{d}r < 0$).  This property is also realized in
local-domain simulations of rapidly rotating convection, where more
rapid rotation leads to enhanced horizontal mixing through vortex
interactions and a resulting decrease in enthalpy transport as
vertical velocities and thermal perturbations become more decorrelated
\citep{Julien_et_al_1996, Brummell_et_al_1996}.  The change of
$\mathrm{d} \bar{S}/\mathrm{d}r$ with rotation rate is shown in
Figure~\ref{fig:ash_structure}$a$ for four of our simulations.  Fixing
the amplitude and structure of $\kappa_0(r)$ across simulations leads
$F_u$ to influence a greater portion of the convection zone at more
rapid rotation, as indicated by the slight growth of $F_u$ in case~G5.
Though this effect becomes stronger as our simulations rotate more
rapidly, at no point in these simulations does $F_u$ transport more
than 10\% of the total luminosity at mid convection zone.

%
%
%

\begin{deluxetable*}{lcccccccc}
\tablecaption{Flux Balances and Energies\label{table:energy balances}}
\tablewidth{0pt}  
\tablehead{\colhead{Case} 
& \colhead{$F_{e,\mathrm{pole}}$\tablenotemark{a}}
& \colhead{$F_{e,\mathrm{eq}}$\tablenotemark{a}}
& \colhead{$F_{k,\mathrm{pole}}$\tablenotemark{a}}
& \colhead{$F_{k,\mathrm{eq}}$\tablenotemark{a}}
& \colhead{CKE\tablenotemark{b}}
& \colhead{DRKE\tablenotemark{b}}
& \colhead{MCKE\tablenotemark{b}}
& \colhead{$\Delta T_\mathrm{max}$\tablenotemark{c}}
}
\startdata
G1  & 1.014 & 1.140 & -0.118 & -0.190
    &$3.28$  & $2.26$  & $0.025$ & 5.50 \\   
G2  & 1.300 & 0.684 & -0.088 & \phn0.046
    &$2.64$  & $13.2$  & $0.015$ & 28.0 \\  
G3  & 1.349 & 0.628 & -0.077 & \phn0.071
    &$2.40$  & $20.5$  & $0.011$ & 53.5 \\
G4  & 1.327 & 0.631 & -0.065 & \phn0.073
    &$2.21$  & $25.5$  & $0.009$ & 78.5 \\ 
G5  & 1.329 & 0.625 & -0.079 & \phn0.078
    &$2.11$  & $30.1$  & $0.007$ & 107  \\ 
G7  & 1.298 & 0.581 & -0.097 & \phn0.080
    &$1.69$  & $38.7$  & $0.005$ & 171  \\ 
G10 & 1.236 & 0.623 & -0.093 & \phn0.090
    &$1.51$  & $47.1$  & $0.003$ & 271  \\[0.25cm]  

G3a & 1.268 & 0.655 & -0.071 & \phn0.068
    &$2.73$  & $27.7$  & $0.012$ & 58.7 \\  
G3b & 1.101 & 0.780 & -0.065 & \phn0.043
    &$3.34$  & $37.9$  & $0.013$ & 62.4 \\  
G5b & 1.172 & 0.668 & -0.047 & \phn0.081 
    &$2.44$  & $57.4$  & $0.008$ & 134  \\[-0.25cm]
\enddata
\tablenotetext{a}{Average convective enthalpy ($F_e$) and kinetic
  energy ($F_k$) fluxes at mid-layer scaled by the solar
  flux, shown for polar (latitudes above $\pm 60^\circ$) and
    equatorial (from $\pm 30^\circ$) regions.}
\tablenotetext{b}{Kinetic energy density relative to
  the rotating coordinate system, for
  convection (CKE), differential rotation (DRKE) and meridional
  circulations (MCKE), averaged over the full shell and over $\sim$ 300 days; units
     are $10^6~\mathrm{erg}\thinspace \mathrm{cm}^{-3}$.}
\tablenotetext{c}{Maximum temperature contrast at $0.96 R_\sol$ in K, typically
  occurring between pole and~$\pm40^\circ$}
\end{deluxetable*}

In all these simulations, the strong correlations between radial
velocities and temperature fluctuations yield the enthalpy flux $F_e$,
which dominates the energy transport at mid convection zone.  
Both warm upflows and cool downflows serve to transport flux out of
the star, and the two carry comparable amounts of flux to one another
in the rapidly rotating simulations. 
In going to the more rapidly rotating simulations, we find that the
average convective enthalpy flux through the polar regions is greater
than that through the equator.  This is shown in
Table~\ref{table:energy balances}.  In case~G1 (at the solar rate)
slightly more flux is transported through the equator than the poles, but
as the rotation rate increases significantly more flux is transported
through the poles than the equator.  This latitudinal variation
becomes somewhat weaker in our more turbulent cases (G3a, G3b and
G5b), but remains present.  Interestingly, cases G7 and G10 follow
similar trends in their equatorial enthalpy transport, despite the
emergence of strong nests of convection at the equator and the
suppression of convection in the rest of that region.

In case~G1, convection transports slightly more than
the solar luminosity.  This over luminosity is balanced by an inward
transport of kinetic energy, which is primarily due to compressible
effects and the transport of $v_r^2$ within strong downflows
that span the convection zone and feel the full density
stratification \citep{Hurlburt_et_al_1986}.  
In the more rapidly rotating cases, the prominent differential rotation
shears apart the convection cells.  The downflows have
lost much of their coherence and only the strongest downflows within
the nests of localized convection survive to span the full
convection zone.  Individual downflows thus feel less density
stratification and compressible effects become less important, leading
to a balance in the transport of $v_r^2$ between the upflows and
downflows.  Instead, as shown in Table~\ref{table:energy balances} the
sense of $F_k$ reverses in the equatorial regions,
becoming dominated by the outward transport of $v_\phi^2$.  The polar
regions remain largely unchanged and the total $F_k$ across spherical surfaces
is nearly zero.

Volume-averaged energy densities for a selection of our simulations
are shown in Table~\ref{table:energy balances}.  At the solar rotation
rate, convective kinetic energy (with kernel
$\frac{1}{2}\bar{\rho}v^{'2}$ and labeled CKE) and the kinetic energy in the
average differential rotation 
($\frac{1}{2}\bar{\rho}\langle v_\phi \rangle^2$, DRKE) are comparable.  As the
rotation rate is increased, DRKE grows strongly and convective energy
decreases slightly, leading DRKE to dominate the total energy budget.
This is true even in our significantly more turbulent solutions.  
The energy in meridional circulations ($\frac{1}{2}\bar{\rho}(\langle
v_r \rangle^2 + \langle v_\theta \rangle^2)$, MCKE) is always small,
and decreases in both magnitude and percentage of the total energy
with more rapid rotation.


\section{Active Nests of Convection}
The emergence of localized nests of convection at higher rotation
rates is a striking feature that calls out for an explanation.  In
many ways, it is quite surprising that convection chooses to be
confined to narrow intervals in longitude, but such states have also
been realized in a number of other dynamical systems.  Generally the
appearance of nests is a challenge to explain in detail, yet the onset
of spatially modulated states which may be their precursor is better
understood. 

\label{sec:patches}
\subsection{Spatially Localized Convection in Other Settings}
The phenomena of spatially localized convection has a rich history,
variously appearing in laboratory experiments and numerical
simulations.  Much interest in confined states of convection was
motivated by the discovery of such states in binary fluid convection
\citep[e.g.,][]{ Anderson&Behringer_1990, Kolodner&Glazier_1990,
Niemela_et_al_1990, Surko_et_al_1991}, where traveling waves of
convection appear via subcritical Hopf bifurcations and near onset are
seen to evolve into traveling patches of convection separated by
regions of nearly quiescent fluid.  From a theoretical perspective,
these confined states near onset are accessible to weakly nonlinear
theory and considerable progress has been made in understanding their nature
\citep[e.g.,][]{Riecke_1992,Barten_et_al_1995b, Batiste_Knobloch_2005,
Batiste_et_al_2006, Burke_Knobloch_2007}.  The confined states found
in binary convection differ from our active nests of convection in
several respects.  The most important is that in binary convection
there is no net vertical transport of solute.  The confined states
instead pump solute horizontally and create regions of stable vertical
stratification in the quiescent regions.  A possibly related phenomena
is that of localized states in magnetoconvection, as studied by
\cite{Blanchflower_1999} and in 3-D by \cite{Blanchflower_Weiss_2002}.
Here single convective cells (called ``convectons'') formed in a
region of initially uniform strong vertical magnetic field by a process of
flux expulsion.  Convection within the localized states was strong and
was entirely suppressed in surrounding medium.  These convectons could
contain several convective cells and were generally stationary, though
some solutions exhibited time dependent behavior.  Recently, progress
has been made addressing these systems in approximate 2-D models
\citep{Dawes_2007}.

Confined states are
also realized in other doubly diffusive systems, as in theoretical
studies of thermosolutal
convection \citep{Knobloch_et_al_1986, Deane_et_al_1987,
  Deane_et_al_1988, Spina_et_al_1998}.  In the latter studies a
variety of traveling convective patches were found, and in these
the convective transport of heat and solute was enhanced
compared to that in the interpatch regions.  In all cases the patches
propagated in the same direction as the individual convective cells,
though more slowly.  Such behavior persisted for long periods of time.  
These localized states occurred well above the onset of convection,
and convection continued in the interpatch regions.  There may also be
analogues in convection within the Earth's atmosphere, where
deep convection in the tropics tends to be organized on global scales
into regions of locally enhanced convection which propagate in a
prograde sense.  These organized convective structures are called the
Madden-Julien Oscillation and appear to have their origin in the
coupling of convective motions with equatorially trapped waves
 \citep[perhaps Rossby or Kelvin waves; see review by][]{Zhang_2005}.    

\begin{figure*}
  \plotone{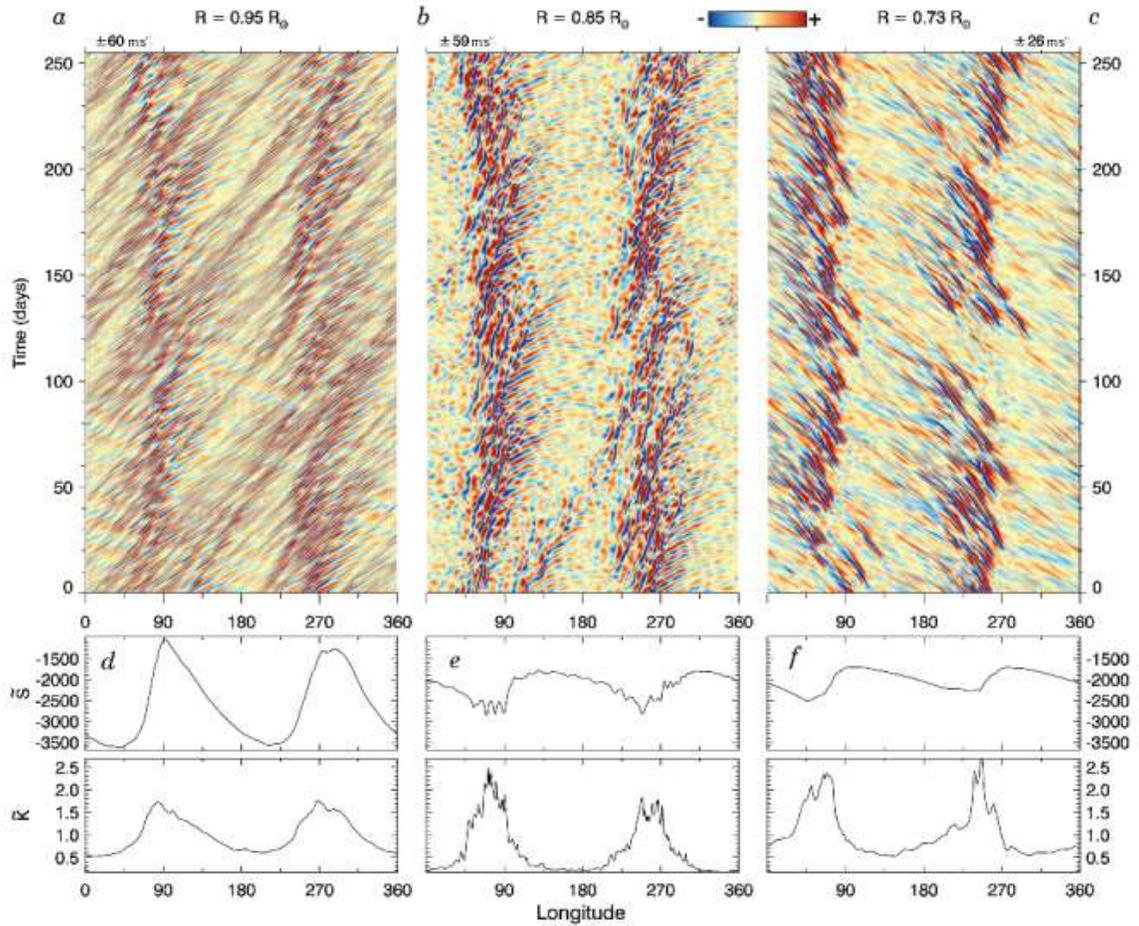}
  \caption{%
  Time-longitude map for case~G5 of radial velocity $v_r$ sampled at three depths 
  ($a$) near top, ($b$) middle and ($c$) bottom of layer.
  These samples are extracted at the equator, using a reference frame
  tracking the nests and starting from a mature state in the simulation.
  Two nests are visible, and individual convective cells appear
  as paired upflows and downflows, which propagate slightly faster than the mean zonal flow that they
  establish and thus pass through the nests.
  ($d,e,f$) Time averages with longitude of entropy fluctuations
  $\widetilde{S}$ and convective kinetic energy density $\widetilde{K}$ in these samples.
  \label{fig:time_longitude_map_turf5_all_3_depths}}
\end{figure*}

The spatially modulated states in our simulations of stellar
convection exist at Rayleigh numbers far above the onset of
convection.  Spatially modulated states in this parameter regime have
also been observed in geophysically motivated 3-D Boussinesq simulations of
convection within a thick, rotating spherical shell
\citep{Grote&Busse_2000, Busse_2002, Busse_Simitev_2005}. In these
studies, spatially localized states emerged at moderate Rayleigh
numbers, involving an equilibrium between the shearing flow of
differential rotation destroying convective eddies and the Reynolds
stresses in the convection driving the differential rotation.  These
effects led to localized states where convection occupied a limited
portion of the domain and the region outside the convective patch was
filled with quiescent streaming flow and almost no radial motions.  In
the quiescent regions the thermal gradients become increasingly
unstable until they are advected back into the patch where they help
sustain the convective eddies.  The patches in these geophysical
simulations moved slowly retrograde and persisted up to Rayleigh
numbers of about $10^6$.  Beyond this point the differential rotation
became so strong that no sustained convection was possible.  Instead
the system began to behave as a relaxation oscillator, with short
bursts of convection temporarily building a strong differential
rotation which then sheared out the convective eddies.  Convective
transport remained suppressed until the shear of differential rotation
decayed viscously, whereupon a new burst of convection would begin the
process anew.  In all cases with localized convection, significant
oscillations were seen in the kinetic energies of both differential
rotation and convection \citep{Grote&Busse_2000}.

Similar states have also been realized in anelastic simulations of
stellar convection in spherical shells for a rotating younger sun with
a much thicker convection zone
\citep{Ballot_et_al_2006,Ballot_et_al_2007}.  The spatially modulated
states found there appear in the equatorial convection for models with
low Prandtl number ($\mathrm{Pr}=0.25$, as in the models of this paper).
Localized states turn into bursty, vacillating convection at large
Taylor numbers ($\mathrm{Ta} \gtrsim 10^9$), much like those in
\cite{Grote&Busse_2000}.  Localized states observed in thick
convective shells (with typical aspect ratios $\chi =
r_\mathrm{bot}/r_\mathrm{top}=0.58$) differ in many important respects
from the states we find in our relatively thin shells of convection
($\chi = 0.76$), most notably in their temporal stability, which we
will next discuss.

\subsection{Properties of the Active Nests}
Our active nests of convection appear first as regions of mildly
enhanced convective amplitude.  As the rotation rate increases, convection
in the equatorial regions gradually becomes more localized (as in Fig.~\ref{fig:ab2_turf}) and eventually is
present only within the active nest.  In some of our systems we observe two
nests or patches in longitude (such as case~G5) and in some only a
single nest (as in case~G10).  Generally,
convection at the highest rotation rates possesses only a single nest, and
for moderate rotation rates the system can alternate between two-nest
states and single-nest states, suggesting that several attractors
exist within the phase space.

To study the temporal evolution of our convective patterns in more
detail, we employ time-longitude maps as shown in
Figure~\ref{fig:time_longitude_map_turf5_all_3_depths}.  Here radial
velocities are sampled at the equator for all longitudes, considering
case~G5 at three depths (near the top, middle and bottom of the
convection zone) and doing so over an interval of 260 days (or about
45 rotation periods).  In these maps, structures propagating in a
prograde fashion relative to the frame of reference are tilted to the
upper-right and patterns propagating in a retrograde sense tilt to the
upper-left.  To construct these maps we have chosen a frame of
reference propagating in a prograde sense relative to the bulk
rotation frame ($\Omega_0=1.3 \times 10^{-5}~\mathrm{rad}~\mathrm{s}^{-1}$ 
or 2070~nHz), with constant relative angular velocity 
$6.75 \times 10^{-7}~\mathrm{rad}~\mathrm{s}^{-1}$
(107~nHz, for a total rotation rate of $1.052~\Omega_0$).  This
corresponds to the propagation rate of the modulated convection
pattern, and thus the nests appear stationary in this frame.  In the
$\Omega_0$ reference frame the nest takes about 108~days to complete
one lap around the equator, thus for the time interval shown in
Figure~\ref{fig:time_longitude_map_turf5_all_3_depths} the nests
travel about $870^\circ$ in longitude and completes about 2.4 circuits
around the equator. The active nests of convection can persist for
extremely long periods of time.  The two nests visible in
Figure~\ref{fig:time_longitude_map_turf5_all_3_depths} remain
coherent for over 5000 days of simulated time.

Individual convective upflows and downflows appear as streaked red and blue regions
respectively.  In the upper convection zone ($0.95 R_\sol$), the
convective cells propagate more rapidly than the nests of convection.
Here they overtake the nests from behind (from smaller longitudes) and
then slowly swim through at a reduced speed.  Within the nests
convective cells collide and interact, and the strongest survive to
emerge from the front of the nest, where they speed up as they then
propagate through the more quiescent region.  
When this occurs, typically a small wave train comprised of 2 to 3
upflow/downflow pairs escapes, and as they enter the quiescent regions
these convective cells speed up to once again outpace the zonal flow.
In the lower convection
zone ($0.73 R_\sol$), the nests propagate more rapidly than the
convective cells, and the individual upflows and downflows appear as
strong retrograde-directed streaks.  Also visible in the lower
convection zone are low-amplitude velocity structures of rapid retrograde
propagation.  They appear to be the weak
equatorward extension of the large-scale (retrograde rotating) polar
patterns evident in Figure~\ref{fig:G5_thermal_structure}.  At
mid-convection zone ($0.85 R_\sol$), nearly all vertical flow is
occurring within the nests of active convection, though the strongest
cells outside the nest in the upper convection zone manage to weakly
print down to this intermediate depth.

\begin{figure}
  \plotone{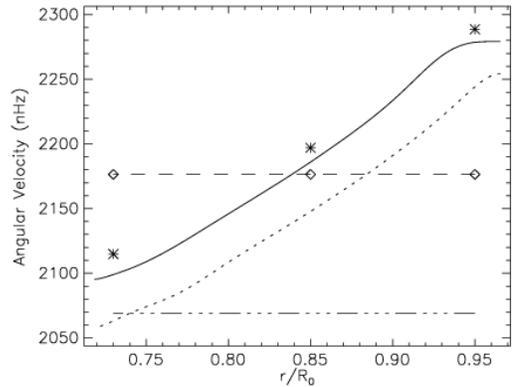}
  \caption{Various angular velocities with radius for features in case~G5.  The mean
  shearing zonal flow of differential rotation $\Omega$ is denoted for all
  depths at the equator (\emph{solid curve}) and at $\pm
  15^\circ$ latitude (\emph{dotted}).  The propagation rate of the
  active nests (\emph{diamonds}) is constant across the 
  entire latitude and radial range where they appear ($\pm 30^\circ$).
  The average velocity of individual convective cells at the equator
  (\emph{asterisks}) is more rapid than the mean zonal flow they
  establish.  Their propagation is faster than the nests
  near the top and slower near the bottom.
  The global rotation rate $\Omega_0$ is 2070 nHz (marked). 
  \label{fig:patch_velocity_cartoon}}
\end{figure}

The typical angular velocities of the differential rotation, the
individual convective cells and the active nests of convection are
shown in Figure~\ref{fig:patch_velocity_cartoon} for case~G5 and
detailed for several cases in Table~\ref{table:angular velocities}.  The nests of
convection propagate at a constant angular velocity at all depths in
the convection zone and over the entire range of latitudes ($\pm
30^\circ$) where they are present.  In contrast, the angular velocity $\Omega$
of the differential rotation varies substantially with radius and
latitude.  At all depths, the individual convective cells propagate
more rapidly than the zonal flow of differential rotation which they drive.  Because
the nests of convection propagate at an intermediate prograde rate,
they experience a head-wind from the differential rotation in the deep
convection zone and a tail-wind near the
surface.  Despite this strong radial shear, the
nests remain coherent across the entire convection zone for long
periods compared with either the lifetime of individual convective
cells (10-30 days), the rotation period of the star (5 days) or
typical thermal and viscous diffusion times ($\tau_\kappa =
910~\mathrm{days}$, $\tau_\nu=3640~\mathrm{days}$, both at mid-depth).
The time for the differential rotation to lap the nests at the
equator is about 112 days near the surface and 143 days at the bottom of the shell.

\begin{deluxetable}{lcccccc}
\tablecaption{Angular Velocities of Various Structures\label{table:angular velocities}}
\tablewidth{0pt}  
\tablehead{\colhead{Case} 
& \colhead{$\Omega_0$}
& \colhead{$\Omega_\mathrm{nest}$}
& \colhead{$\Omega_{\mathrm{eq},\mathrm{top}}$}
& \colhead{$\Omega_{\mathrm{eq},\mathrm{bot}}$}
& \colhead{$\Omega_{c,\mathrm{top}}$}
& \colhead{$\Omega_{c,\mathrm{bot}}$}
}
\startdata
G1  & \phn2.60 & ---   & 0.341 & 0.060 & 0.535 & 0.180 \\
G3  & \phn7.80 & 0.511 & 1.094 & 0.220 & 1.176 & 0.330 \\ 
G5  & 13.00    & 0.675 & 1.319 & 0.197 & 1.381 & 0.288 \\ 
G10 & 26.00    & 0.830 & 1.497 & 0.140 & 1.623 & 0.280 \\[0.25cm]  

G3a & \phn7.80 & 0.690 & 1.295 & 0.265 & 1.421 & 0.373 \\ 
G3b & 13.00    & 0.670 & 1.555 & 0.250 & 1.590 & 0.413 \\
G5b & 13.00    & 0.750 & 1.778 & 0.246 & 1.800 & 0.430 \\[-0.25cm]
\enddata
\tablecomments{All angular velocities in
  $\mu\mathrm{rad}\thinspace\mathrm{s}^{-1}$ and, except for the frame
  rate $\Omega_0$, are given relative to $\Omega_0$.
  The differential rotation at the equator $\Omega_\mathrm{eq}$ is measured at
  $0.95R_\sol$ (\emph{top}) and $0.73R_\sol$ (\emph{bot}).   The mean propagation rate of
  individual convective structures $\Omega_c$ is measured at the same
  depths in time-longitude maps of $v_r$ and has a typical variance of 
  $\pm 0.04~\mu\mathrm{rad}\thinspace\mathrm{s}^{-1}$ .} 
\end{deluxetable}

\begin{figure}
  \plotone{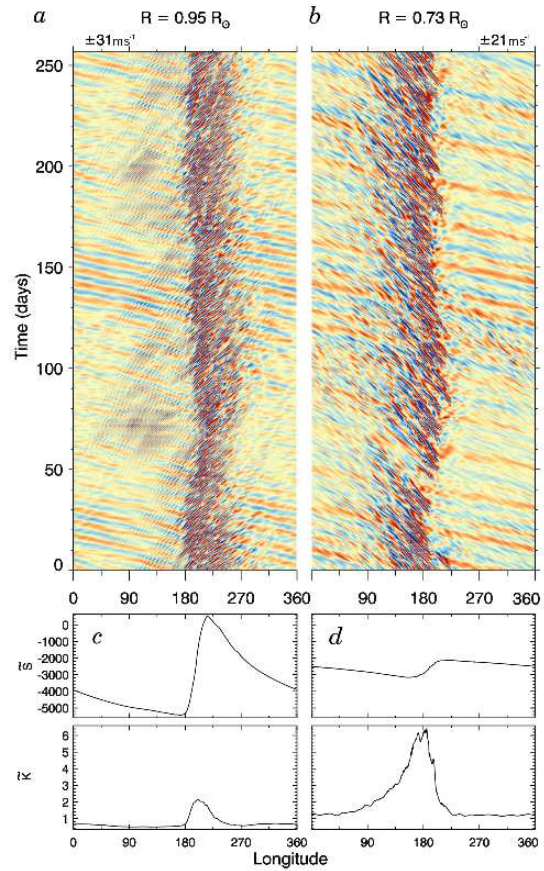}
  \caption{%
  As in Fig.~\ref{fig:time_longitude_map_turf5_all_3_depths},
  time-longitude map for case~G10 of radial velocity ($a$) near top
  and ($b$) near bottom of layer.
  Here a single nest is realized, with individual
  convective cells almost entirely confined to the nest, though they
  continue to propagate within it.
  ($c,d$) Time averages $\widetilde{S}$ and $\widetilde{K}$ constructed in the
  same reference frame.
  \label{fig:time_longitude_map_turf10_2_depths}}
\end{figure}

At the higher rotation rates, single stable nests of convection
dominate the equatorial region.  Time-longitude maps are shown in
Figure~\ref{fig:time_longitude_map_turf10_2_depths} for the equatorial
radial velocity at two depths in case~G10, our most rapidly rotating simulation.
Within the nest, convection remains vigorous, with the strongest
downflow networks still spanning the entire depth of the convection
zone.  This nest again propagates in a prograde sense relative to
$\Omega_0$, with constant relative angular velocity $8.3\times
10^{-7}~\mathrm{rad}~\mathrm{s}^{-1}$ (132~nHz, for a total rotation
rate of $1.032~\Omega_0$).  Over the interval shown the nest completes
almost 3 circuits of the equator relative to the $\Omega_0$ reference
frame.  Individual convective cells continue to swim through the nest,
moving more rapidly near the surface and more slowly in the deep
convection zone.  The very strong radial shear prevents all but the
strongest of downflows from spanning the full convection zone.  In the
region outside the nest, convection is strongly suppressed and the
main features are the weak flows associated with the retrograde
propagating polar pattern.  In the upper convection zone, occasional
weak convective disturbances appear upstream of the nest.  As these
fluctuations enter the nest they grow in amplitude.

Our nests of active convection may owe their existence to a
competition between the shearing action of differential rotation
acting on the individual convective eddies and Reynolds stresses
within the convection helping to maintain the strong zonal flows.  Unlike
the systems studied by \cite{Grote&Busse_2000} and
\cite{Ballot_et_al_2007}, our patchy convection is not accompanied by
relaxation oscillation behavior or large exchanges between the kinetic
energy in convection and in the differential rotation.  Rather, our
nests are not bursty in time and instead persist for long periods in
quasi-steady states.  This is true even for our most rapidly rotating
case considered here (G10) and at much higher turbulent
driving (G5b), though these simulations exist at Taylor
numbers below the threshold suggested by \cite{Ballot_et_al_2007}
($\mathrm{Ta} \gtrsim 10^9$).  Coupling between equatorially trapped waves and
convection may also have a role, but the contribution
of this coupling to spatial localization has been difficult to elucidate.

\subsection{Detailed Structure of an Active Nest of Convection}

\begin{figure}
  \plotone{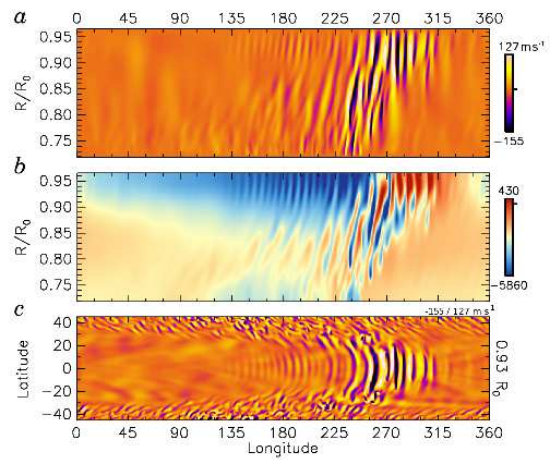}
  \caption{%
    Snapshot of nest structure in case~G10, at relative day 221 of
    Fig.~\ref{fig:time_longitude_map_turf10_2_depths}.  ($a$) Radial
    velocities in an equatorial cut shown for all longitudes and
    radii.  
    ($b$) Companion entropy fluctuations $S$ about
    their spherical means $\bar{S}$ (in cgs units), with high entropy
    fluid in red tones.
    ($c$) Radial velocity in the equatorial region shown near the top of layer.
    \label{fig:G10_patch}}
\end{figure}

We focus here on the structure of the nests so evident in
Figures~\ref{fig:time_longitude_map_turf5_all_3_depths} and
\ref{fig:time_longitude_map_turf10_2_depths} for cases~G5 and~G10, but
devote particular attention to the single nest realized in the latter
as a representative case.  The active nests extend throughout the
depth of the convection zone.  This is illustrated most clearly in
Figure~\ref{fig:G10_patch}$a$, showing the vertical profile
of radial velocities with longitude in a cut at the equator.
Convection is broken into multiple cells, one set above $0.9 R_\sol$
and another below $0.8 R_\sol$.  Only within the nest do strong downflows
span the convection zone.  The nest is embedded in a region of strong
latitudinal and radial zonal shear (as shown by the tilted nature of
its structure), yet the pattern propagates at a constant angular
velocity at all depths and latitudes.  At the surface and near the
equator the nest propagates more slowly than the zonal wind,
whereas at the base of the convection zone it propagates more rapidly.
Individual convective structures swim still more rapidly and both
enter and exit the region of enhanced convection.  As such, the nest
experiences a strong retrograde flow at the base and a strong prograde
flow near the surface.  

The thermal structure of the nests is revealed by their entropy
fluctuations, as shown in Figure~\ref{fig:G10_patch}$b$.  In the upper
convection zone, the mean zonal flow advects low entropy fluid into
the nests from the left side.  This fluid is then swept away by
intermittent downflows and replaced by higher
entropy fluid from below.  In the lower convection zone, higher entropy
fluid is swept into the nest from the right and lower entropy exits to
the left.  At mid convection zone, regions outside the nest remain
convectively unstable, but only weak radial motions are driven here.

\begin{figure}
  \plotone{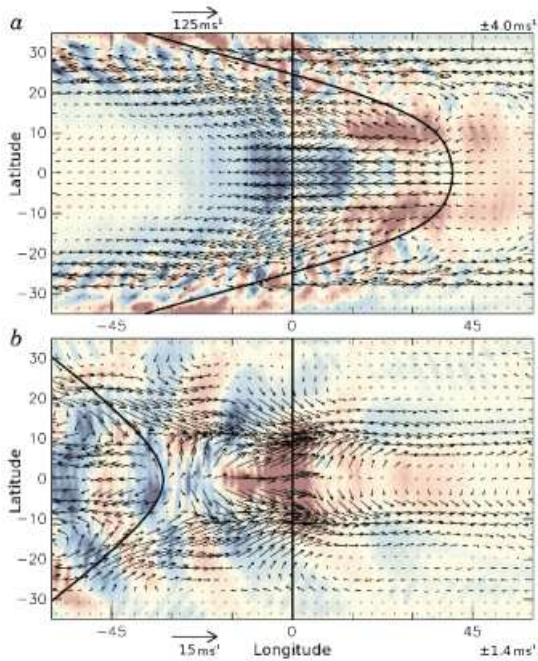}
  \caption{Mean circulations associated with the nest in case~G10.  
  ($a$) Time averaged radial (colors) and horizontal flows (arrows)
  near the top and ($b$) near the bottom of the shell, in a reference
  frame tracking the nest.  The strong zonal flow of differential rotation has
  been subtracted at each latitude and 
  is shown by the solid curves (both scaled by
  125~\ms arrow length, with zero velocity relative to the
  nest indicated by the vertical line).
  Within the nest strong eddy currents partially decelerate
  the flow, while outside the streaming mean zonal flow dominates. 
  \label{fig:G10_tracked_nest}}
\end{figure}

The mean longitudinal structure of the nests can be assessed by
forming temporal averages of various properties in a frame co-rotating
with the nests of convection (these averages denoted by a tilde).
This is done at the equator for entropy $\widetilde{S}$ and 
convective kinetic energy density $\widetilde{K}$ (with same form as CKE) 
for case~G5 in Figure~\ref{fig:time_longitude_map_turf5_all_3_depths}$d-f$ and for 
case~G10 in Figure~\ref{fig:time_longitude_map_turf10_2_depths}$c,d$.
There are distinctive differences between the leading (to the right)
and trailing (to the left) portions of the profiles, with $\widetilde{S}$ showing
a gradual rise and steeper drop in going to decreasing longitudes
(from right to left).  The envelope of $\widetilde{S}$ is largely similar in form
at the top and bottom of the convection zone.  In contrast $\widetilde{K}$,
which traces the fluctuating velocities of convection, changes form
with depth, being skewed in the direction of the mean zonal flow.  At
the top of the convection zone this sense of skew is toward the
right, and at the bottom it is toward the left.

These nests must have gradual circulations associated with them, though these
are a challenge to discern as they are much weaker than the strong
mean zonal flows.  We can get some sense of them by averaging the
vertical and horizontal flows in the surroundings of a nest while
tracking it over long time intervals.  Shown in
Figure~\ref{fig:G10_tracked_nest} are the slow circulations associated
with the nest in case~G10 near the top and bottom of the convection zone.  
These flows have been averaged over a period of 615~days starting
at the beginning of the interval shown in
Figure~\ref{fig:time_longitude_map_turf10_2_depths}.  
The flows suggest a systematic zonal circulation from fore to aft in the upper
convection zone and toward higher latitudes and diverging from the nest
at depth, though to achieve this we have subtracted the fast mean
zonal flow which varies with latitude as shown.  
The mean upflows and downflows in the nest are quite weak, with amplitudes
of a few \ms, as compared with the convection which can have
peak velocities of a few hundred \ms.
Test particles released in the flow would not simply circulate
according to these mean circulations and would
be instead swept along by the strong mean zonal flow and of course by
the vigorous convection cells that propagate through the nest.  Some
test particles would encounter strong downflows and would be swept
down through the nest to the bottom of the convection zone while
others would likely be carried horizontally out the nest and remain at
a similar depth.
Yet Figure~\ref{fig:G10_tracked_nest} indicates that a weak large-scale
circulation is realized, and this may serve to slowly pump fluid
through the system.  These mean circulatory flows may also serve to inform
analytic models of such nests of convection.


\section{Conclusion}
\label{sec:conclusion}

When stars like our sun are young they rotate much more rapidly than
the present sun.  In these stars rotation must strongly
influence the convective motions and may lead to stronger global-scale
dynamo action.
We have explored here the effects of more rapid rotation on global-scale
convection in simulations of stars like our sun.  The mean zonal flows of
differential rotation become much stronger with more rapid rotation,
scaling as $\Delta \Omega \propto \Omega_0^{0.3}$ or as
$\Delta \Omega/\Omega_\mathrm{eq} \propto \Omega_0^{-0.6}$.  In striking
contrast, the meridional circulations become much weaker with more
rapid rotation, and the energy contained in them drops approximately
as $\Omega_0^{-0.9}$.  Accompanying the growing differential rotation
is a significant latitudinal temperature contrast, with amplitudes of
$100~\mathrm{K}$ or higher in the most rapidly rotating cases.  The maximum
temperature contrast near the surface occurs between the hot poles and
the cool mid latitudes at about $\pm 40^\circ$.  If this latitudinal
temperature contrast prints through the vigorous convection at the
stellar surface, it may appear as an observable latitudinal variation
in intensity.  The thermal contrast would presumably persist for long
periods compared to stellar activity, offering a way to disentangle
this intensity signature from that caused by spots of magnetism at the
stellar poles.

These simulations are entirely hydrodynamic and this provides a major
caveat to our findings on the scaling of differential rotation 
and latitudinal temperature contrast with rotation rate $\Omega_0$.
Prior MHD simulations of stellar convection have demonstrated that in
some parameter regimes strong dynamo-generated magnetic fields can
react back strongly on the differential rotation, acting to lessen
angular velocity contrasts or largely eliminate them
\citep[e.g.,][]{Brun_et_al_2005, Featherstone_et_al_2007,Browning_2008}.  
It is unclear whether the scaling trends identified here for
differential rotation as a function of $\Omega_0$ will survive in the
presence of dynamo action and magnetic fields.  Likewise, magnetic
fields may lessen the strong temperature contrasts realized here,
doing so through their feedback on the convective flows and energy transport.
We expect that the weaker meridional circulations may be less affected
by magnetic feedbacks, and thus the prediction that meridional
circulations lessen in energy and amplitude with more rapid rotation
may be of greater significance though harder to confirm
observationally.  Weaker meridional circulations in more rapidly
rotating stars will have a strong impact on many theories of stellar
dynamo action, including the Babcock-Leighton flux-transport model favored for
solar-type stars. 
We have begun simulations to explore the dynamo action realized at
various rotation rates, and its impact upon the flows described here.
Preliminary results appear in \cite{Brown_et_al_2007c} and more
detailed results will be forthcoming shortly.

A striking feature of these simulations is the emergence of a 
pattern of strongly modulated convection in the equatorial regions.  These
nests of active convection are regions of enhanced convective vigor
and transport which propagate at rates distinct from either the mean
zonal flows of differential rotation or the individual convective
cells.  In the most rapidly rotating systems, such as case~G10,
convection at the equator is entirely dominated by motions inside the
nest with only very weak radial motions present in the regions
outside the nest.  Though their impact on the convection is most
obvious in the rapidly rotating limit, we find some evidence for weak
modulation even in our more slowly rotating cases.

All of our simulations stop short of the turbulent stellar surface,
and it is thus difficult to estimate how these nests of active
convection may affect stellar observations in detail.  The convective velocities
associated with the nests are small compared to the nearly supersonic
flows in stellar granulation, and in the sun such global-scale
convective structures have evaded direct detection despite intensive
searches throughout the near-surface layers by helioseismology.  The
extremely localized nests found in our most rapidly rotating cases may
however influence the thermal stratification and thus convective vigor
in the near surface regions, as most of the flux at the equator is
transported through a narrow range of longitudes.  These nests may act
as traveling hot spots with enhanced convection even in the surface
layers where the higher emerging flux escapes the system.

These spatially localized states of convection may also have some bearing
on the active longitudes of magnetic activity observed in the sun, if
the enhanced pummeling of the tachocline by the convection within the
nest preferentially destabilizes magnetic structures within the
tachocline that then rise to the surface.  Initial dynamo simulations
indicate that nests of convection can coexist with magnetism in
portions of parameter space.  Thus their strongest signature is
likely to emerge in magnetic stars, where magnetic fields threading
the bulk of the convection zone may be concentrated in the nests and 
mimic giant, propagating starspots which survive for very long epochs
in time.  If the nests lead to active
longitudes of enhanced magnetic activity in rapidly rotating stars, we
might expect these long-lived magnetic structures to propagate at a
rate different from the stellar rotation rate as measured at the
surface or from the stellar differential rotation.

We recognize that our simulations remain separated
by many orders of magnitude from the parameter space of real stellar
convection.  As such, we must be cautious with our interpretations of
the overall dynamics.  However, we have found that these nests of
convection are a robust feature over a range of parameters and that
they are able to persist as entities for as long as we could pursue their
modelling.  Thus one should be prepared to consider the possibility of
their presence also in real stellar convection zones, where they may
appear as long-lived propagating features.

\acknowledgements
This research is supported by NASA through Heliophysics Theory
Program grants NNG05G124G and NNX08AI57G, with additional support for
Brown through the NASA GSRP program by award number
NNG05GN08H.  Browning was supported by a NSF Astronomy and Astrophysics
postdoctoral fellowship AST 05-02413.
The simulations were carried out with NSF PACI support of
PSC, SDSC and NCSA, and by NASA support at Project Columbia.

\bibliographystyle{apj}


\begin{thebibliography}{69}
\expandafter\ifx\csname natexlab\endcsname\relax\def\natexlab#1{#1}\fi

\bibitem[{{Anderson} \& {Behringer}(1990)}]{Anderson&Behringer_1990}
{Anderson}, K.E., \& {Behringer}, R.P. 1990, Phys. Lett. A, 145, 323

\bibitem[{{Ballot} {et~al.}(2006){Ballot}, {Brun}, \&
  {Turck-Chi{\`e}ze}}]{Ballot_et_al_2006}
{Ballot}, J., {Brun}, A.~S., \& {Turck-Chi{\`e}ze}, S. 2006, in Beyond the
  Spherical Sun: A New Era of Helio- and Asteroseismology, ed. K. Fletcher,
  \& M. Thompson, (ESA SP-624; Noordwijk: ESA), 108.1

\bibitem[{{Ballot} {et~al.}(2007){Ballot}, {Brun}, \&
  {Turck-Chi{\`e}ze}}]{Ballot_et_al_2007}
{Ballot}, J., {Brun}, A.S., \& {Turck-Chi{\`e}ze}, S. 2007, \apj, 669, 1190

\bibitem[{{Barnes} {et~al.}(2005){Barnes}, {Cameron}, {Donati}, {James},
  {Marsden}, \& {Petit}}]{Barnes_et_al_2005}
{Barnes}, J.R., {Cameron}, A.C., {Donati}, J.-F., {James}, D.J., {Marsden},
  S.C., \& {Petit}, P. 2005, \mnras, 357, L1

\bibitem[{{Barten} {et~al.}(1995){Barten}, {L{\"u}cke}, {Kamps}, \&
  {Schmitz}}]{Barten_et_al_1995b}
{Barten}, W., {L{\"u}cke}, M., {Kamps}, M., \& {Schmitz}, R. 1995, \pre, 51,
  5662

\bibitem[{{Batiste} \& {Knobloch}(2005)}]{Batiste_Knobloch_2005}
{Batiste}, O., \& {Knobloch}, E. 2005, Phys. Rev. Lett., 95, 244501

\bibitem[{{Batiste} {et~al.}(2006){Batiste}, {Knobloch}, {Alonso}, \&
  {Mercader}}]{Batiste_et_al_2006}
{Batiste}, O., {Knobloch}, E., {Alonso}, A., \& {Mercader}, I. 2006, J.
  Fluid Mech., 560, 149

\bibitem[{{Blanchflower}(1999)}]{Blanchflower_1999}
{Blanchflower}, S. 1999, Phys. Lett. A, 261, 74

\bibitem[{{Blanchflower} \& {Weiss}(2002)}]{Blanchflower_Weiss_2002}
{Blanchflower}, S., \& {Weiss}, N.O. 2002, Phys. Lett. A, 294, 297

\bibitem[{{Brown} {et~al.}(2007){Brown}, {Browning}, {Brun}, {Miesch},
  {Nelson}, \& {Toomre}}]{Brown_et_al_2007c}
{Brown}, B.P., {Browning}, M.K., {Brun}, A.S., {Miesch}, M.S., {Nelson},
  N.J., \& {Toomre}, J. 2007, in Unsolved Problems in Stellar Physics, ed. 
  R.J. Stancliffe et al., Amer. Inst. Phys., CP-948, 271.

\bibitem[{{Brown} {et~al.}(2004){Brown}, {Browning}, {Brun}, \&
  {Toomre}}]{Brown_et_al_2004}
{Brown}, B.P., {Browning}, M.K., {Brun}, A.S., \& {Toomre}, J. 2004, in 
  Helio- and Asteroseismology: Towards a Golden Future, ed. D. {Danesy}, 
  (ESA SP-559; Noordwijk: ESA), 341

\bibitem[{{Browning}(2008)}]{Browning_2008}
{Browning}, M.K. 2008, \apj, 676, 1262

\bibitem[{{Browning} {et~al.}(2004){Browning}, {Brun}, \&
  {Toomre}}]{Browning_et_al_2004}
{Browning}, M.K., {Brun}, A.S., \& {Toomre}, J. 2004, \apj, 601, 512

\bibitem[{{Browning} {et~al.}(2006){Browning}, {Miesch}, {Brun}, \&
  {Toomre}}]{Browning_et_al_2006}
{Browning}, M.K., {Miesch}, M.S., {Brun}, A.S., \& {Toomre}, J. 2006, \apjl,
  648, L157

\bibitem[{{Brummell} {et~al.}(1996){Brummell}, {Hurlburt}, \&
  {Toomre}}]{Brummell_et_al_1996}
{Brummell}, N.H., {Hurlburt}, N.E., \& {Toomre}, J. 1996, \apj, 473, 494

\bibitem[{{Brummell} {et~al.}(1998){Brummell}, {Hurlburt}, \&
  {Toomre}}]{Brummell_et_al_1998}
---. 1998, \apj, 493, 955

\bibitem[{{Brun} {et~al.}(2002){Brun}, {Antia}, {Chitre}, \&
  {Zahn}}]{Brun_et_al_2002}
{Brun}, A.S., {Antia}, H.M., {Chitre}, S.M., \& {Zahn}, J.-P. 2002, \aap,
  391, 725

\bibitem[{{Brun} {et~al.}(2005){Brun}, {Browning}, \&
  {Toomre}}]{Brun_et_al_2005}
{Brun}, A.S., {Browning}, M.K., \& {Toomre}, J. 2005, \apj, 629, 461

\bibitem[{{Brun} {et~al.}(2004){Brun}, {Miesch}, \& {Toomre}}]{Brun_et_al_2004}
{Brun}, A.S., {Miesch}, M.S., \& {Toomre}, J. 2004, \apj, 614, 1073

\bibitem[{{Brun} \& {Toomre}(2002)}]{Brun&Toomre_2002}
{Brun}, A.S., \& {Toomre}, J. 2002, \apj, 570, 865

\bibitem[{{Burke} \& {Knobloch}(2007)}]{Burke_Knobloch_2007}
{Burke}, J., \& {Knobloch}, E. 2007, Phys. Lett. A, 360, 681

\bibitem[{{Busse}(1970)}]{Busse_1970}
{Busse}, F.H. 1970, J. Fluid Mech., 44, 441

\bibitem[{{Busse}(2002)}]{Busse_2002}
---. 2002, Phys. Fluids, 14, 1301

\bibitem[{{Busse} \& {Simitev}(2005)}]{Busse_Simitev_2005}
{Busse}, F.H., \& {Simitev}, R. 2005, Astron. Nachr., 326, 231

\bibitem[{{Chandrasekhar}(1961)}]{Chandrasekhar_1961}
{Chandrasekhar}, S. 1961, {Hydrodynamic and Hydromagnetic Stability}
   (Oxford: Clarendon)

\bibitem[{{Charbonneau}(2005)}]{Charbonneau_2005}
{Charbonneau}, P. 2005, Living Rev. Sol. Phys, 2, 2, 
   http://www.livingreviews.org/lrsp-2005-2
  

\bibitem[{{Charbonneau} \& {Saar}(2001)}]{Charbonneau&Saar_2001}
{Charbonneau}, P., \& {Saar}, S.H. 2001, in Magnetic Fields Across the 
  Hertzsprung-Russell Diagram, ed. G.~{Mathys}, S.K. {Solanki}, \& D.T. 
  {Wickramasinghe}, ASP Conf. Ser., Vol. 248, 189

\bibitem[{{Clune} {et~al.}(1999){Clune}, {Elliott}, {Glatzmaier}, {Miesch}, \&
  {Toomre}}]{Clune_et_al_1999}
{Clune}, T.L., {Elliott}, J.R., {Glatzmaier}, G.A., {Miesch}, M.S., \&
  {Toomre}, J. 1999, Parallel Comp., 25, 361

\bibitem[{{Dawes}(2007)}]{Dawes_2007}
{Dawes}, J.H.P. 2007, J. Fluid Mech., 570, 385

\bibitem[{{Deane} {et~al.}(1987){Deane}, {Toomre}, \&
  {Knobloch}}]{Deane_et_al_1987}
{Deane}, A.E., {Toomre}, J., \& {Knobloch}, E. 1987, \pra, 36, 2862

\bibitem[{{Deane} {et~al.}(1988){Deane}, {Toomre}, \&
  {Knobloch}}]{Deane_et_al_1988}
---. 1988, \pra, 37, 1817

\bibitem[{{Dikpati} {et~al.}(2001){Dikpati}, {Saar}, {Brummell}, \&
  {Charbonneau}}]{Dikpati_et_al_2001}
{Dikpati}, M., {Saar}, S.H., {Brummell}, N., \& {Charbonneau}, P. 2001, 
  in Magnetic Fields Across the Hertzsprung-Russell Diagram, eds. G.~{Mathys}, 
  S.K. {Solanki}, \& D.T. {Wickramasinghe}, ASP Conf. Ser., Vol. 248, 235

\bibitem[{{Donahue} {et~al.}(1996){Donahue}, {Saar}, \&
  {Baliunas}}]{Donahue_et_al_1996}
{Donahue}, R.A., {Saar}, S.H., \& {Baliunas}, S.L. 1996, \apj, 466, 384

\bibitem[{{Donati} {et~al.}(2003){Donati}, {Cameron}, {Semel}, {Hussain},
  {Petit}, {Carter}, {Marsden}, {Mengel}, {L{\'o}pez Ariste}, {Jeffers}, \&
  {Rees}}]{Donati_et_al_2003}
{Donati}, J.-F., {Cameron}, A.C., {Semel}, M., {Hussain}, G.A.J., {Petit},
  P., {Carter}, B.D., {Marsden}, S.C., {Mengel}, M., {L{\'o}pez Ariste}, A.,
  {Jeffers}, S.V., \& {Rees}, D.E. 2003, \mnras, 345, 1145

\bibitem[{{Dormy} {et~al.}(2004){Dormy}, {Soward}, {Jones}, {Jault}, \&
  {Cardin}}]{Dormy_et_al_2004}
{Dormy}, E., {Soward}, A.M., {Jones}, C.A., {Jault}, D., \& {Cardin}, P.
  2004, J. Fluid Mech., 501, 43

\bibitem[{{Featherstone} {et~al.}(2007){Featherstone}, {Browning}, {Brun}, \&
  {Toomre}}]{Featherstone_et_al_2007}
{Featherstone}, N., {Browning}, M.K., {Brun}, A.S., \& {Toomre}, J. 2007,
  Astron. Nachr., 328, 1126

\bibitem[{{Gilman}(1975)}]{Gilman_1975}
{Gilman}, P.~A. 1975, J. Atmos. Sci., 32, 1331

\bibitem[{{Gilman}(1977)}]{Gilman_1977}
---. 1977, Geophys. Astrophys. Fluid Dynam., 8, 93

\bibitem[{{Gilman}(1979)}]{Gilman_1979}
---. 1979, \apj, 231, 284

\bibitem[{{Gilman} \& {Glatzmaier}(1981)}]{Gilman&Glatzmaier_1981}
{Gilman}, P.A., \& {Glatzmaier}, G.A. 1981, \apjs, 45, 335

\bibitem[{{Glatzmaier} \& {Gilman}(1981)}]{Glatzmaier&Gilman_1981}
{Glatzmaier}, G.A., \& {Gilman}, P.A. 1981, \apjs, 45, 351

\bibitem[{{Grote} \& {Busse}(2000)}]{Grote&Busse_2000}
{Grote}, E., \& {Busse}, F.H. 2000, Fluid Dynam. Res., 28, 349

\bibitem[{{Hurlburt} {et~al.}(1986){Hurlburt}, {Toomre}, \&
  {Massaguer}}]{Hurlburt_et_al_1986}
{Hurlburt}, N.E., {Toomre}, J., \& {Massaguer}, J.M. 1986, \apj, 311, 563

\bibitem[{{Julien} {et~al.}(1996){Julien}, {Legg}, {McWilliams}, \&
  {Werne}}]{Julien_et_al_1996}
{Julien}, K., {Legg}, S., {McWilliams}, J., \& {Werne}, J. 1996, J.
  Fluid Mech., 322, 243

\bibitem[{{Knobloch} {et~al.}(1986){Knobloch}, {Moore}, {Toomre}, \&
  {Weiss}}]{Knobloch_et_al_1986}
{Knobloch}, E., {Moore}, D.R., {Toomre}, J., \& {Weiss}, N.O. 1986, J.
  Fluid Mech., 166, 409

\bibitem[{{Kolodner} \& {Glazier}(1990)}]{Kolodner&Glazier_1990}
{Kolodner}, P., \& {Glazier}, J.A. 1990, \pra, 42, 7504

\bibitem[{{K{\"u}ker} \&
  {R{\"u}diger}(2005{\natexlab{a}})}]{Kuker&Rudiger_2005_A&A}
{K{\"u}ker}, M., \& {R{\"u}diger}, G. 2005{\natexlab{a}}, \aap, 433, 1023

\bibitem[{{K{\"u}ker} \&
  {R{\"u}diger}(2005{\natexlab{b}})}]{Kuker&Rudiger_2005_AN}
---. 2005{\natexlab{b}}, Astron. Nachr., 326, 265

\bibitem[{{K{\"u}ker} \& {Stix}(2001)}]{Kuker&Stix_2001}
{K{\"u}ker}, M., \& {Stix}, M. 2001, \aap, 366, 668

\bibitem[{{Miesch}(1998)}]{Miesch_thesis}
{Miesch}, M.S. 1998, Ph.D. thesis, University of Colorado, Boulder

\bibitem[{{Miesch}(2005)}]{Miesch_2005}
---. 2005, Living Rev. Sol. Phys, 2, 1, http://www.livingreviews.org/lrsp-2005-1


\bibitem[{{Miesch} {et~al.}(2008){Miesch}, {Brun}, {DeRosa}, \&
  {Toomre}}]{Miesch_et_al_2008}
{Miesch}, M.S., {Brun}, A.S., {DeRosa}, M.L., \& {Toomre}, J. 2008, \apj,
  673, 557

\bibitem[{{Miesch} {et~al.}(2006){Miesch}, {Brun}, \&
  {Toomre}}]{Miesch_et_al_2006}
{Miesch}, M.S., {Brun}, A.S., \& {Toomre}, J. 2006, \apj, 641, 618

\bibitem[{{Miesch} {et~al.}(2000){Miesch}, {Elliott}, {Toomre}, {Clune},
  {Glatzmaier}, \& {Gilman}}]{Miesch_et_al_2000}
{Miesch}, M.~S., {Elliott}, J.R., {Toomre}, J., {Clune}, T.L., {Glatzmaier},
  G.A., \& {Gilman}, P.A. 2000, \apj, 532, 593

\bibitem[{{Niemela} {et~al.}(1990){Niemela}, {Ahlers}, \&
  {Cannell}}]{Niemela_et_al_1990}
{Niemela}, J.J., {Ahlers}, G., \& {Cannell}, D.S. 1990, Phys. Rev.
  Lett., 64, 1365

\bibitem[{{Noyes} {et~al.}(1984){Noyes}, {Hartmann}, {Baliunas}, {Duncan}, \&
  {Vaughan}}]{Noyes_et_al_1984a}
{Noyes}, R.W., {Hartmann}, L.W., {Baliunas}, S.L., {Duncan}, D.K., \&
  {Vaughan}, A.H. 1984, \apj, 279, 763

\bibitem[{{Patten} \& {Simon}(1996)}]{Patten&Simon_1996}
{Patten}, B.M., \& {Simon}, T. 1996, \apjs, 106, 489

\bibitem[{{Pedlosky}(1982)}]{Pedlosky_1987}
{Pedlosky}, J. 1982, {Geophysical Fluid Dynamics} (New York: Springer-Verlag)

\bibitem[{{Reiners}(2006)}]{Reiners_2006}
{Reiners}, A. 2006, \aap, 446, 267

\bibitem[{{Reiners} \& {Schmitt}(2003)}]{Reiners&Schmitt_2003}
{Reiners}, A., \& {Schmitt}, J.H.M.M. 2003, \aap, 398, 647

\bibitem[{{Rempel}(2005)}]{Rempel_2005}
{Rempel}, M. 2005, \apj, 622, 1320

\bibitem[{{Rempel}(2008)}]{Rempel_2008}
---. 2008, in {Helioseismology, Asteroseismology and MHD Connections},
           J. Physics Conf. Ser. (IoP), in press

\bibitem[{{Riecke}(1992)}]{Riecke_1992}
{Riecke}, H. 1992, Phys. Rev. Lett., 68, 301

\bibitem[{{R{\"u}diger} {et~al.}(1998){R{\"u}diger}, {von Rekowski}, {Donahue},
  \& {Baliunas}}]{Rudiger_et_al_1998}
{R{\"u}diger}, G., {von Rekowski}, B., {Donahue}, R.A., \& {Baliunas}, S.L.
  1998, \apj, 494, 691

\bibitem[{{Spina} {et~al.}(1998){Spina}, {Toomre}, \&
  {Knobloch}}]{Spina_et_al_1998}
{Spina}, A., {Toomre}, J., \& {Knobloch}, E. 1998, \pre, 57, 524

\bibitem[{{Surko} {et~al.}(1991){Surko}, {Ohlsen}, {Yamamoto}, \&
  {Kolodner}}]{Surko_et_al_1991}
{Surko}, C.M., {Ohlsen}, D.R., {Yamamoto}, S.Y., \& {Kolodner}, P. 1991,
  \pra, 43, 7101

\bibitem[{{Thompson} {et~al.}(2003){Thompson}, {Christensen-Dalsgaard},
  {Miesch}, \& {Toomre}}]{Thompson_et_al_2003}
{Thompson}, M.~J., {Christensen-Dalsgaard}, J., {Miesch}, M.S., \& {Toomre},
  J. 2003, \araa, 41, 599

\bibitem[{{Walker} {et~al.}(2007){Walker}, {Croll}, {Kuschnig}, {Walker},
  {Rucinski}, {Matthews}, {Guenther}, {Moffat}, {Sasselov}, \&
  {Weiss}}]{Walker_et_al_2007}
{Walker}, G.A.H., {Croll}, B., {Kuschnig}, R., {Walker}, A., {Rucinski},
  S.M., {Matthews}, J.M., {Guenther}, D.B., {Moffat}, A.F.J., {Sasselov},
  D., \& {Weiss}, W.W. 2007, \apj, 659, 1611

\bibitem[{{Zahn}(1992)}]{Zahn_1992}
{Zahn}, J.-P. 1992, \aap, 265, 115

\bibitem[{{Zhang}(2005)}]{Zhang_2005}
{Zhang}, C. 2005, Rev. Geophys., 43, G2003


\end{thebibliography}

\end{document}